\documentclass[fleqn,usenatbib]{mnras}

\usepackage{newtxtext,newtxmath}
\usepackage[T1]{fontenc}
\usepackage{ae,aecompl}

\usepackage{graphicx}	% Including figure files
\usepackage{amsmath}	% Advanced maths commands
\usepackage{hyperref}
\usepackage{natbib}
\usepackage[caption=false]{subfig}
\usepackage{ragged2e}
\usepackage{placeins}
\usepackage{bigints} %To be able to write bigger integral symbols
\usepackage{physics}
\usepackage{xcolor}
\usepackage{cleveref}
\usepackage[export]{adjustbox}
\usepackage{longtable}
\usepackage{xtab}

\usepackage{scalerel,tikz}
\usetikzlibrary{svg.path}
\definecolor{orcidlogocol}{HTML}{A6CE39}
\tikzset{orcidlogo/.pic={
 \fill[orcidlogocol] svg{M256,128c0,70.7-57.3,128-128,128C57.3,256,0,198.7,0,128C0,57.3,57.3,0,128,0C198.7,0,256,57.3,256,128z};
 \fill[white] svg{M86.3,186.2H70.9V79.1h15.4v48.4V186.2z}
 svg{M108.9,79.1h41.6c39.6,0,57,28.3,57,53.6c0,27.5-21.5,53.6-56.8,53.6h-41.8V79.1z M124.3,172.4h24.5c34.9,0,42.9-26.5,42.9-39.7c0-21.5-13.7-39.7-43.7-39.7h-23.7V172.4z}
 svg{M88.7,56.8c0,5.5-4.5,10.1-10.1,10.1c-5.6,0-10.1-4.6-10.1-10.1c0-5.6,4.5-10.1,10.1-10.1C84.2,46.7,88.7,51.3,88.7,56.8z};
}}
\newcommand\orcidicon[1]{\href{https://orcid.org/#1}{\mbox{\scalerel*{
\begin{tikzpicture}[yscale=-1,transform shape]
\pic{orcidlogo};
\end{tikzpicture}
}{|}}}}

\def\mean#1{\left< #1 \right>}

\defcitealias{Mohapatra2020}{MFS20}
\defcitealias{Shi2019}{SZ19}
\title[Density fluctuations in the turbulent ICM]{Turbulent density and pressure fluctuations in the stratified intracluster medium}

\author[Mohapatra, Federrath \& Sharma]{
Rajsekhar Mohapatra$^{\orcidicon{0000-0002-1600-7552}\,1}$\thanks{E-mail: rajsekhar.mohapatra@anu.edu.au (RM)},
Christoph Federrath$^{\orcidicon{0000-0002-0706-2306}\,1,2}$\thanks{E-mail: christoph.federrath@anu.edu.au (CF)} and 
Prateek Sharma$^{\orcidicon{0000-0003-2635-4643}\,3}$\thanks{E-mail: prateek@iisc.ac.in (PS)}
\\
$^{1}$Research School of Astronomy and Astrophysics, Australian National University, Canberra, ACT~2611, Australia\\
$^{2}$Australian Research Council Centre of Excellence in All Sky Astrophysics (ASTRO3D), Canberra, ACT~2611, Australia\\
$^{3}$Department of Physics, Indian Institute of Science, Bangalore, KA 560012, India
}

\date{Accepted XXX. Received YYY; in original form ZZZ}

\pubyear{2020}

\begin{document}
\label{firstpage}
\pagerange{\pageref{firstpage}--\pageref{lastpage}}
\maketitle

\begin{abstract}
Turbulent gas motions are observed in the intracluster medium (ICM). The ICM is density-stratified, with the gas density being highest at the centre of the cluster and decreasing radially outwards. As a result of this, Kolmogorov (homogeneous, isotropic) turbulence theory does not apply to the ICM. The gas motions are instead explained by anisotropic stratified turbulence, with the stratification quantified by the perpendicular Froude number ($\mathrm{Fr}_\perp$). These turbulent motions are associated with density and pressure fluctuations, which manifest as perturbations in X-ray surface brightness maps of the ICM and as thermal Sunyaev-Zeldovich effect (SZ) fluctuations, respectively. In order to advance our understanding of the relations between these fluctuations and the turbulent gas velocities, we have conducted 100 high-resolution hydrodynamic simulations of stratified turbulence ($256^2\times 384$---$1024^2\times1536$ resolution elements), in which we scan the parameter space of subsonic rms Mach number ($\mathcal{M}$), $\mathrm{Fr}_\perp$, and the ratio of entropy and pressure scale heights ($R_{PS}=H_P/H_S$), relevant to the ICM. We develop a new scaling relation between the standard deviation of logarithmic density fluctuations ($\sigma_s$, where $s=\ln(\rho/\mean{\rho})$), $\mathcal{M}$, and $\mathrm{Fr}_{\perp}$, which covers both the strongly stratified ($\mathrm{Fr}_{\perp}\ll1$) and weakly stratified ($\mathrm{Fr}_{\perp}\gg 1$) turbulence regimes:~$\sigma_s^2=\ln\left(1+b^2\mathcal{M}^4+0.10/(\mathrm{Fr}_\perp+0.25/\sqrt{\mathrm{Fr}_\perp})^2\mathcal{M}^2R_{PS}\right)$, where $b\sim1/3$ for solenoidal turbulence driving studied here. We further find that logarithmic pressure fluctuations $\sigma_{(\ln{P}/\mean{P})}$ are independent of stratification and scale according to the relation $\sigma_{(\ln{\bar{P}})}^2=\ln\left(1+b^2\gamma^2\mathcal{M}^4\right)$, where $\bar{P}=P/\mean{P}$ and $\gamma$ is the adiabatic index of the gas. We have tested these scaling relations to be valid over the parameter ranges $\mathcal{M} = 0.01$--$0.40$, $\mathrm{Fr}_\perp = 0.04$--$10.0$, and $R_{PS} = 0.33$--$2.33$.

\end{abstract}


%%%%%%%%%%%%%%%%%%%%%%%%%%%%%%%%%%%%%%%%%%%%%%%%%%

%%%%%%%%%%%%%%%%% BODY OF PAPER %%%%%%%%%%%%%%%%%%

\section{Introduction}\label{sec:introduction}

Turbulence and buoyancy are concurrent in several geophysical and astrophysical systems -- the physics of stratified turbulence governs ocean currents and atmospheric turbulence on the earth and other planets, radiative and convective zones in the atmospheres of the sun and other stars, gas motions in hot gaseous haloes of galaxies (the circumgalactic medium or CGM), galaxy groups and clusters (intracluster medium or ICM)  \citep{Stein1967,Goldreich1977ApJ,Loewenstein1990MNRAS,Sarazin1992ApJ,Rudie2012ApJ,Parmentier2013,Skoutnev2020arXiv}.
Here we focus on stratified turbulence relevant to the ICM. 

ICM refers to the gas that pervades the region between galaxies in a cluster. It is mostly composed of the hot X-ray emitting gas, with temperatures ranging from $10^7$--$10^8\,$K, although a filamentary colder phase has also been detected in many clusters \citep{Cowie1983ApJ,McDonald2010ApJ,Simionescu2018MNRAS,Vantyghem2019ApJ,Olivares2019A&A}. It is moderately stratified (with Richardson number $\mathrm{Ri}\lesssim10$ or Froude number $\mathrm{Fr}_\perp\gtrsim0.1$, refer to figure~1 in \citealt{Mohapatra2020}) and the gas is in rough hydrostatic equilibrium with the gravitational profile set by the dark matter halo. Turbulence in the ICM plays an important role in the gas dynamics and evolution. Turbulent energy dissipation on viscous scales and subsequent heating of the ICM, turbulent mixing of hot and cold phases of gas \citep{Kim2003ApJ,banerjee2014turbulence,Hillel2020ApJ} may also play a key role in the gas thermodynamics, by preventing the runaway cooling of the ICM core \citep{zhuravleva2014turbulent}. Anisotropy in turbulent eddies may be used to probe the orientation of ICM magnetic fields \citep{Hu2020ApJ}. In cluster outskirts, estimating the turbulent pressure support of the gas is important %in the hydrostatic mass bias problem 
to get an unbiased estimate of the halo mass that is required for cosmology with clusters \citep{schuecker2004,George2009,Bautz2009,Cavaliere2011,Nelson2014ApJ,Biffi2016ApJ,Angelinelli2020MNRAS}. 

However, direct measurements of the ICM turbulent gas velocities \citep{hitomi2016} are still a few years away (XRISM\footnote{\url{https://global.jaxa.jp/projects/sas/xrism/}} and Athena\footnote{\url{https://www.the-athena-x-ray-observatory.eu/}}), after the early mission end of the Hitomi satellite. Recently, observers have relied on several indirect methods to estimate gas velocities in the hot phase, such as relating X-ray surface brightness fluctuations to turbulent velocity fluctuations \citep{zhuravleva2014turbulent,Zhuravleva2015MNRAS,Zhuravleva2019NatAs}, relating Sunyaev-Zeldovich effect (SZ) observations to turbulent pressure fluctuations \citep{Zeldovich1969,khatri2016,Mroczkowski2019}, measuring cold-gas velocities (e.g., using the  H$\alpha$ line) and relating this to the hot phase velocity \citep{Li2020ApJ}. \cite{Simionescu2019SSRv} provide a detailed review of all these different methods of measuring gas velocities in the ICM. It is important to develop accurate and robust scaling relations between the ICM hot-gas velocities and these observables. 

Many recent theoretical studies have focused on the scaling of density fluctuations with the rms Mach number ($\mathcal{M}$) \citep{gaspari2014,zhuravleva2014relation,nolan2015,Mohapatra2019,Shi2019,Grete2020ApJ,Mohapatra2020}. Some of these studies have ignored the effect of gravitational stratification or lack a detailed parameter scan of the range of $\mathrm{Fr}_\perp$ and $\mathcal{M}$ relevant for the ICM. The relation between density fluctuations and turbulent velocities is also seen to depend on the equation of state of the gas \citep{Federrath2015MNRAS} and the adiabatic index \citep{nolan2015}, and whether gas cooling is included \citep{Mohapatra2019,Grete2020ApJ}. 

Several fluid mechanics studies \citep{bolgiano1959,Bolgiano1962,Carnevale2001,Lindborg2006,Brethouwer2008,Herring2013,Kumar2014,Rorai2014PhRvE,Feraco2018arXiv,Alam2019} discuss the theory of stratified turbulence in the context of planetary atmospheres and oceans. In these studies, turbulence is driven perpendicular to the direction of gravity, whereas turbulence in the ICM is driven more isotropically by active galactic nuclei (AGN) jets and galaxy mergers \citep{Churazov2002MNRAS,Churazov2003,Omma2004MNRAS}. They also mainly focus on the scaling of velocity and passive scalar spectra, intermittency
 and velocity anisotropy in the strong stratification limit. 

In our previous study \citep{Mohapatra2020} (hereafter referred to as \citetalias{Mohapatra2020}), we performed stratified turbulence simulations with a fixed $\mathcal{M}$ and scanned the parameter space of weakly and moderately stratified turbulence. We also proposed a scaling relation between density fluctuations, Richardson number ($\mathrm{Ri}$) and $\mathcal{M}$ for this regime. We found that density fluctuations also depended on a third parameter, namely the ratio between pressure and entropy scale heights ($R_{PS}=H_P/H_S$). We found that for $\mathrm{Ri}\lesssim10$ (only moderately stratified), these numbers are sensitive to the turbulent driving length scale $L$, as $\mathrm{Ri}\propto L^2$. Here we scan the parameter space of these three parameters: $\mathcal{M}$, the transverse Froude number $\mathrm{Fr}_\perp$ ($\mathrm{Fr}_\perp\approx1/\sqrt{\mathrm{Ri}}$ for $\mathrm{Fr}_\perp \gtrsim1$) and $R_{PS}$ through 100 simulations, extending into the strongly stratified regime ($\mathrm{Fr}_\perp\ll1$). %relevant for the ICM. 

The paper is organised as follows. In \cref{sec:Methods} we briefly describe our setup and methods, present our results and their interpretations in \cref{sec:results-discussion}, compare our results with the literature and discuss the caveats of our work in \cref{sec:caveats-future}, and conclude in \cref{sec:Conclusion}.

% === Methods ===
\section{Methods}\label{sec:Methods}
\subsection{Model equations}\label{subsec:ModEq}
We model the ICM as a fluid using compressible Euler equations and ideal gas equation of state. We implement gravity and turbulent forcing as additional source terms in the momentum and energy equations.
We solve the following equations:
\begin{subequations}
	\begin{align}
	\label{eq:Euler1}
	&\frac{\partial\rho}{\partial t}+\nabla\cdot (\rho \mathbf{v})=0,\\
	\label{eq:Euler2}
	&\frac{\partial(\rho\mathbf{v})}{\partial t}+\nabla\cdot (\rho \mathbf{v}\otimes \mathbf{v})+\nabla P=\rho\mathbf{F}+\rho \mathbf{g},\\
	\label{eq:Euler3}
	&\frac{\partial E}{\partial t}+\nabla\cdot ((E+P)\mathbf{v})=\rho\mathbf{F}\cdot\mathbf{v}+\rho (\mathbf{v}\cdot\nabla)\Phi,
	\end{align}
\end{subequations}
where $\rho$ is the gas mass density, $\mathbf{v}$ is the velocity, $P=\rho k_B T/(\mu m_p)$ is the pressure (we use the ideal gas equation of state), $\mathbf{F}$ is the turbulent acceleration that we apply, $\Phi$ is the gravitational potential, $\mathbf{g}=-\nabla\Phi$ is the acceleration due to gravity, $E=\rho v^2/2 + P/(\gamma-1)$ 
is the sum of kinetic and internal energy densities, $\mu$ is the mean particle weight, $m_p$ is the proton mass, $k_B$ is the Boltzmann constant, $T$ is the temperature, and $\gamma=5/3$ is the adiabatic index. 
\subsection{Setup}\label{subsec:Setup}
We choose $-\hat{\mathbf{z}}$ to be the direction of the gravitational field, and pressure and density to have scale heights $H_P$ and $H_{\rho}$, respectively. Thus, the initial pressure and density profiles are given by
 \begin{subequations}
 	\begin{align}
 	\label{eq:Presinitial}
 	&P(t=0)=P_0\exp(-\frac{z}{H_P}) \text{ and}\\
 	\label{eq:Densinitial}
 	&\rho(t=0)=\rho_0\exp(-\frac{z}{H_{\rho}})\text{, respectively.} 	
 	\end{align}
 \end{subequations}
We work with dimensionless units and choose $\rho_0=1$ and $P_0=0.6$, so that $c_{s,0}=\sqrt{\gamma P_0/\rho_0}=1$. Since we start with the gas in hydrostatic equilibrium, the initial density, pressure, and $g$, are related by 
 \begin{subequations}
\begin{equation}\label{eq:grav}
 	\frac{\mathrm{d}P}{dz}=-\rho g. 
\end{equation}
 Hence $g$ is set as
	\begin{align}
		\label{eq:grav1}
		g&=\frac{P_0}{\rho_0 H_P}\exp(-z\left[\frac{1}{H_P}-\frac{1}{H_{\rho}}\right]).
	\end{align} 
 \end{subequations}
This equilibrium is convectively stable if $\mathrm{d}\ln S/\mathrm{dz}>0$, where 
\begin{equation}
S=\frac{P}{\rho^{\gamma}}\text{ is the pseudo-entropy.}
\end{equation} This gives us the condition for the entropy scale height $H_S$ ($\equiv 1/[\mathrm{d} \ln S/\mathrm{d}z]$), given by
\begin{equation}\label{eq:H_S}
	\frac{1}{H_S}=\frac{\gamma}{H_{\rho}}-\frac{1}{H_P}>0.
\end{equation}
This condition is satisfied for all our simulations, which locally mimic the stably stratified ICM.

\subsection{Important stratified turbulence parameters}\label{subsec:turb-params}
When a parcel of gas in a stably stratified medium is displaced from its original position, it oscillates with a frequency $N$ defined as the Brunt-V\"{a}is\"{a}l\"{a} (BV) frequency, given by 
\begin{subequations}
\begin{equation}\label{eq:BVoscilations}
    N^2=\frac{g}{\gamma} \frac{d}{dz} \ln \left (\frac{P}{\rho^\gamma} \right).
\end{equation}
We define the turbulent time scale as $\ell_\perp/v_\perp$, where $\ell_\perp$ is the integral scale, defined as 
\begin{align}
    & \ell_{\perp}=2\pi\frac{\int k_\perp^{-1} E_{k_\perp} \mathrm{d}k_\perp}{\int E_{k_\perp} \mathrm{d}k_\perp} \label{l_perp},\\
    & v_{\perp}=\left\langle\frac{1}{2}{\mathbf{v}}_{\perp}^{2}\right\rangle^{1 / 2}.\label{eq:v_perp}
\end{align}   
Here $E_{k_\perp}$ is the velocity power spectrum perpendicular to the direction of $\mathbf{g}$. 
Here $\mathbf{v_\perp}=(v_x,v_y,0)$ denotes the components of the velocity field perpendicular to the direction of gravity.
The perpendicular Froude number $\mathrm{Fr}_\perp$ is the ratio of these two time scales \citep[see e.g., chapter~14 in][]{davidson_2013}, given as 
\begin{equation}
    \mathrm{Fr}_\perp=\frac{v_\perp}{N\ell_\perp}.\label{eq:Fr_perp}
\end{equation}
We can use the approximation $\ell_\perp\approx L_{\text{driv}}$, where $L_{\text{driv}}$ is the driving length scale of the turbulence. 
The parallel Froude number $\mathrm{Fr}_\parallel$ is defined as
\begin{align}
    & \mathrm{Fr}_\parallel = \frac{v_\perp}{Nl_\parallel}\text{, where}\label{eq:Fr_parallel}\\
    & \ell_{\parallel}=2 \pi\frac{\int k_\parallel^{-1} E_{k_\parallel} \mathrm{d}k_\parallel}{\int E_{k_\parallel} \mathrm{d}k_\parallel} \label{l_parallel},
\end{align} 
with $l_\parallel$ being the integral scale parallel to the direction of gravity and $E_{k_\parallel}$ is the velocity power spectrum parallel to the direction of $\mathbf{g}$. Notice that the transverse velocity ($v_\perp$, and not $v_z$) is used in the above expressions as $v_z$ peaks at scales smaller than $l_\perp$. In this study, we use $\mathrm{Fr}_\perp$ to quantify the relative strength of stratification to turbulence\footnote{In \citetalias{Mohapatra2020}, we used the Richardson number ($\mathrm{Ri}$) to quantify the strength of stratification, which is the ratio of buoyancy and turbulence terms in the momentum equation. 
For weakly and moderately stratified turbulence (the parameter regime that we scanned in \citetalias{Mohapatra2020}), we used $\mathrm{Ri}\approx N^2L_{\text{driv}}^2/v_{L_{\text{driv}}}^2$. The assumption of isotropic eddies breaks down for strongly stratified turbulence, which is why we now parametrise the stratification in terms of the transverse Froude number (Fr$_\perp$).}.
The scale-dependent Froude numbers $\Tilde{\mathrm{Fr}}_\perp(\Tilde{\ell}_\perp)$ and $\Tilde{\mathrm{Fr}}_\parallel(\Tilde{\ell}_\parallel)$, for an eddy of size  ($\Tilde{\ell}_\perp$ and $\Tilde{\ell}_\parallel$), and perpendicular velocity $\Tilde{v}_\perp$ are defined as 
\begin{align}
    & \Tilde{\mathrm{Fr}}_\perp(\Tilde{\ell}_\perp)=\frac{\Tilde{v}_\perp(\Tilde{\ell}_\perp)}{N\Tilde{\ell}_\perp},\label{eq:scale_Fr_perp}\\
    & \Tilde{\mathrm{Fr}}_\parallel(\Tilde{\ell}_\parallel)= \frac{\Tilde{v}_\perp(\Tilde{\ell}_\perp)}{N\Tilde{\ell}_\parallel}.\label{eq:scale_Fr_parallel}
\end{align} 
$\mathrm{Fr}_\perp\ll1$ indicates strong stratification \citep[in this regime, $\mathrm{Fr}_\parallel\approx1$, see][]{Billant2001PhF} and $\mathrm{Fr}_\perp\gg1$ denotes weak stratification. Strongly stratified turbulence transitions into weakly stratified turbulence at the Ozmidov length scale $\ell_O$, which is defined as the scale on which the relative strengths of buoyancy and turbulence terms become equal. Since $\Tilde{\mathrm{Fr}}_\perp(\ell_O)=1$,
\begin{equation}
    \ell_O=\sqrt{\epsilon_K/N^3}, \label{eq:l_Ozmidov} 
\end{equation}
where $\epsilon_K=v_\perp^3/\ell_\perp$ is the kinetic energy transfer rate and is assumed to be a constant just as in conventional turbulence.
For weak and moderately stratified turbulence ($\mathrm{Fr}_\perp\gtrsim1$), $\mathrm{Fr}_\perp\approx(1/\mathrm{Ri})^{0.5}$. 

Similar to \citetalias{Mohapatra2020}, density, pressure and velocity are normalised to construct dimensionless variables, such that $\bar{\rho}=\rho/\mean{\rho(z)}$, $\bar{P}=P/\mean{P(z)}$ and $\mathcal{M}=\mean{v/c_s}_{\mathrm{rms}}$, where $\mean{\rho(z)}$ and $\mean{P(z)}$ are the average density and pressure at a $z-$ slice, respectively, $v$ is the amplitude of velocity, and $c_s \equiv (\gamma \langle P \rangle /\langle \rho \rangle)^{1/2}$ is the local speed of sound. 
The potential energy per unit mass is defined as 
\begin{equation}
    E_{u_b}=u_b^2/2=\frac{P}{2\rho}\frac{\gamma\delta\bar{\rho}^2}{R_{PS}}\label{eq:buoy_energy} \text{,}
\end{equation}
\end{subequations}
where $u_b=g\delta\bar{\rho}/N$ is the strength of density fluctuations expressed in velocity units, and
\begin{equation} \label{eq:rps}
R_{PS}=H_P/H_S
\end{equation}
is the ratio of pressure and entropy scale heights. The kinetic energy is defined as $E_{u}=\delta v^2/2$, where $\delta v$ is the magnitude of the fluctuating velocity. 

\subsection{Numerical methods}\label{subsec:numerical_methods}
We evolve the Euler equations (\ref{eq:Euler1} to \ref{eq:Euler3}) using the hydrodynamic version of the HLL5R Riemann solver \citep{Bouchut2007,Bouchut2010,Waagan2011} in a modified version of the FLASH code \citep{Fryxell2000,Dubey2008}, version 4. Our setup is the same as in \citetalias{Mohapatra2020} -- we use a uniformly spaced 3D grid, with a box size $L_x=L_y=1$ and $L_z=1.5$, centred at $(0,0,0)$. 

\subsubsection{Boundary conditions}\label{subsubsec:BC}
We use periodic boundary conditions for all variables (density, pressure and velocity) along the $x$ and $y$ directions. Along the $z$ direction, we use reflective boundary conditions for velocity and Dirichlet boundary conditions for pressure and density with the guard cells filled according to \cref{eq:Presinitial} and (\ref{eq:Densinitial}), respectively. In \citetalias{Mohapatra2020}, we used reflective boundary conditions along the $z$ direction, which led to a hydrostatic instability at the $z$ direction boundaries (due to an inverted pressure gradient at these boundaries). This instability was stronger for strongly stratified turbulence simulations and led to anomalous sound waves moving in the $z$ direction, starting from the boundaries \citep[also seen in][hereafter \citetalias{Shi2019}]{Shi2019}. The Dirichlet boundary conditions used here let us avoid this instability and allow us to extend our study to include the strongly stratified turbulence limit (down to $\mathrm{Fr}_\perp\sim0.05$).

We restrict our analyses to a cube of size $1$ centred at $(0,0,0)$, with boundaries at $(\pm0.5,\pm0.5,\pm0.5)$, to avoid any other anomalous effects near the $z$ direction boundaries. We run our simulations with shallow density profile ($H_\rho>1$) on grids with $256^2\times 384$ resolution elements, and the simulations with steep density profile ($H_\rho\leq1$) or simulations with high rms Mach number $(\mathcal{M}\approx0.4)$ on grids $512^2\times 768$ resolution elements. We also run four strongly stratified simulations at resolution $1024^2 \times 1536$ for numerical convergence checks.

\subsubsection{Turbulent forcing}\label{subsubsec:Turb_forcing}
We follow the same spectral forcing method as in \citetalias{Mohapatra2020}. We use the stochastic Ornstein-Uhlenbeck (OU) process to model the turbulent acceleration $\mathbf{F}$ with a finite autocorrelation time scale $t_{\mathrm{turb}}$ \citep{eswaran1988examination,schmidt2006numerical,federrath2010}. We inject power as a parabolic function of $|k|$, for $1\leq|k|\leq3$ (note that we have dropped the wavenumber unit $2\pi/L$). The power peaks at $|k|_{\text{inj}}=2$, i.e., $L_{\text{driv}}=L/2$. For $k\geq3$, turbulence develops self-consistently. We set $t_{\mathrm{turb}}=L_{\text{driv}}/\sigma_v$, where $\sigma_v$ is the standard deviation of the velocity on $L_{\text{driv}}$. Our driving is solenoidal (zero divergence). For further details of the forcing method, refer to section~2.7 of \citetalias{Mohapatra2020} and section~2.1 in \cite{federrath2010}. 

We also use the same window function, $w(z)$, on the acceleration field, as in \citetalias{Mohapatra2020}, such that $\mathbf{F}$ decays to zero near the boundaries in the $z$ direction, given by
\begin{align}
w(z) =
\begin{cases}
1, & \text{ for }\abs{z}<0.625,\nonumber\\
       \exp(-\abs{\abs{z}-0.625}/0.125), & \text{ for }\abs{z}>0.625.
\end{cases}
\end{align}
Note that $w(z)=1$ inside the analysis box ($|x|,|y|,|z|\leq0.5$). So it only serves to exponentially decrease the acceleration amplitudes close to the $z$ boundaries.

\subsection{List of Simulation models}\label{subsec:list_of_models}
We have conducted 100 simulations, scanning $\mathrm{Fr}_\perp$ between $0.05$--$12.0$, $\mathcal{M}$ between $0.01$--$0.4$ and $R_{PS}$ between $0.33$--$2.33$, covering the parameter range relevant for the stratified ICM. We have three input parameters ($H_P$, $H_\rho$ and the acceleration field) that we vary in our various simulations to scan the range of interest in $\mathcal{M}$, Fr$_\perp$, and $R_{PS}$.%, which we parametrise by $\mathcal{M}_\text{bin}$. 
The bin Mach number,  $\mathcal{M}_{\text{bin}}$, indicates the rms Mach number that the acceleration field would produce in a homogeneous isotropic setup. We use these bins to separate our runs with different Mach numbers. The actual $\mathcal{M}$ can be slightly different from $\mathcal{M}_{\text{bin}}$, especially in models with steep temperature profiles ($H_\rho,\,H_P\lesssim0.25$), as $c_s$ can vary by an order of magnitude with height. By definition, $R_{PS}$ depends only on $H_P/H_\rho$ (c.f., \Cref{eq:rps}). $\mathrm{Fr}_\perp$ depends on all three input parameters, roughly $\mathrm{Fr}_\perp\propto \mathcal{M}_{\text{bin}}H_P/\sqrt{R_{PS}}$ (as $u_\perp \propto \mathcal{M}_{\text{bin}}$ and $N^2\propto1/[H_PH_S]$).

In \cref{tab:sim_params}, we provide a compressed list of the simulation models, grouped under their common $\mathcal{M}_{\text{bin}}$ and $R_{PS}$. We indicate the range of $H_\rho$, the resolution, $\mathcal{M}$, $\mathrm{Fr}_\perp$, and $\sigma_s^2$ ($\sigma_s$ is the standard deviation of $s=\ln(\rho/\mean{\rho})$ in the analysis box) for each of these groups. In \cref{tab:sim_list}, we have expanded each grouping and list all of these parameters individually for each of the 100~simulations.

\begin{table*}
	\centering
	\caption{Simulation parameters for different runs}
	\label{tab:sim_params}
	\resizebox{\textwidth}{!}{
		\begin{tabular}{lccccc} 
			\hline
			Label & $H_{\rho}$ range & Resolution  & Actual $\mathcal{M}$ range& $\mathrm{Fr}_{\perp}$ range  &   $\sigma_s^2$ range \\
			(1) & (2) & (3) & (4) & (5) & (6)\\
			\hline
			$\mathcal{M}0.01R_{PS}0.67$ & $8.0$---$1.0$  & $512^2\times768$& $0.0087$---$0.010$ & $0.34$---$0.04$  & $6.4\times10^{-6}$---$1.2\times10^{-5}$\\
			\hline
			$\mathcal{M}0.05R_{PS}0.33$ & $17.3$---$0.55$  & $256^2\times384$---$512^2 \times 768$  & $0.047$---$0.053$ & $9.5$---$0.16$  & $1.7\times10^{-6}$---$1.4\times10^{-4}$\\
			$\mathcal{M}0.05R_{PS}0.67$ & $21.3$---$0.28$  & $256^2\times384$---$1024^2\times1536$  & $0.048$---$0.057$ & $11.0$---$0.07$  & $2.1\times10^{-6}$---$2.5\times10^{-4}$\\
			$\mathcal{M}0.05R_{PS}1.0$  & $21.9$---$0.2$   & $256^2\times384$---$512^2 \times 768$  & $0.048$---$0.06$  & $11.0$---$0.05$  & $2.5\times10^{-6}$---$3.7\times10^{-4}$\\
			$\mathcal{M}0.05R_{PS}1.5$  & $17.8$---$0.2$   & $256^2\times384$---$512^2 \times 768$  & $0.047$---$0.063$ & $8.80$---$0.06$  & $4.2\times10^{-6}$---$5.8\times10^{-4}$\\
			$\mathcal{M}0.05R_{PS}2.33$ & $20.0$---$0.2$   & $256^2\times384$---$1024^2\times1536$  & $0.046$---$0.064$ & $10.0$---$0.06$  & $4.1\times10^{-6}$---$9.6\times10^{-4}$\\
			\hline
			$\mathcal{M}0.10R_{PS}0.33$ & $8.7$---$0.273$  & $256^2\times384$---$512^2 \times 768$  & $0.096$---$0.104$ & $9.5$---$0.16$  & $1.7\times10^{-5}$---$6.0\times10^{-4}$\\
			$\mathcal{M}0.10R_{PS}0.67$ & $9.6$---$0.22$   & $256^2\times384$---$512^2 \times 768$  & $0.094$---$0.114$  & $9.8$---$0.12$  &  $2.1\times10^{-5}$---$1.1\times10^{-3}$\\
			$\mathcal{M}0.10R_{PS}1.0$  & $11.2$---$0.22$  & $256^2\times384$---$512^2 \times 768$  & $0.094$---$0.114$  & $11.0$---$0.12$  & $2.1\times10^{-5}$---$1.4\times10^{-3}$\\
			$\mathcal{M}0.10R_{PS}1.5$  & $10.6$---$0.22$  & $256^2\times384$---$512^2 \times 768$  & $0.096$---$0.114$  & $11.0$---$0.12$  & $2.4\times10^{-5}$---$2.5\times10^{-3}$\\
			$\mathcal{M}0.10R_{PS}2.33$ & $10.0$---$0.2$   & $256^2\times384$---$512^2 \times 768$  & $0.097$---$0.114$  & $11.0$---$0.11$  & $2.8\times10^{-5}$---$4.7\times10^{-3}$\\
			\hline
			$\mathcal{M}0.25R_{PS}0.33$ & $3.5$---$0.125$  & $256^2\times384$---$512^2 \times 768$  & $0.21$---$0.25$   & $8.9$---$0.16$  & $4.3\times10^{-4}$---$4.2\times10^{-3}$\\
			$\mathcal{M}0.25R_{PS}0.67$ & $6.4$---$0.12$   & $256^2\times384$---$1024^2\times1536$  & $0.23$---$0.27$   & $11.0$---$0.18$  & $5.0\times10^{-4}$---$8.5\times10^{-3}$\\
			$\mathcal{M}0.25R_{PS}1.0$  & $7.0$---$0.10$   & $256^2\times384$---$512^2 \times 768$  & $0.23$---$0.36$   & $12.0$---$0.19$  & $5.1\times10^{-4}$---$2.3\times10^{-2}$\\
			$\mathcal{M}0.25R_{PS}1.5$  & $6.4$---$0.10$   & $256^2\times384$---$512^2 \times 768$  & $0.23$---$0.45$   & $11.0$---$0.22$   & $5.9\times10^{-4}$---$6.7\times10^{-2}$\\
			$\mathcal{M}0.25R_{PS}2.33$ & $6.0$---$0.1$    & $256^2\times384$---$1024^2\times1536$  & $0.23$---$0.55$   & $11.0$---$0.29$  & $6.4\times10^{-4}$---$1.4\times10^{-1}$\\
			\hline
			$\mathcal{M}0.40R_{PS}0.67$ & $1.8$---$0.15$   & $512^2\times768$                       & $0.36$---$0.40$   & $7.4$---$0.33$  & $2.4\times10^{-3}$---$3.2\times10^{-2}$\\
			$\mathcal{M}0.40R_{PS}1.0$  & $1.75$---$0.17$  & $512^2\times768$                       & $0.37$---$0.44$   & $7.8$---$0.42$  & $2.6\times10^{-3}$---$4.5\times10^{-2}$\\
			$\mathcal{M}0.40R_{PS}1.5$  & $1.4$---$0.14$   & $512^2\times768$                       & $0.37$---$0.54$   & $6.8$---$0.47$  & $2.9\times10^{-3}$---$9.0\times10^{-2}$\\
			$\mathcal{M}0.40R_{PS}2.33$ & $1.3$---$0.12$   & $512^2\times768$                       & $0.37$---$0.58$   & $7.3$---$0.71$  & $3.0\times10^{-3}$---$1.0\times10^{-1}$\\
		\hline
			
	\end{tabular}}
	\justifying \\ \begin{footnotesize} Notes: Column 1 shows the simulation name. The numbers following `$\mathcal{M}$' and `$R_{PS}$' are the binned Mach number ($\mathcal{M}_\text{bin}$) and the ratio of pressure to entropy scale heights $H_P/H_S$ in the simulations, respectively. In columns~2 and 3 we list the $H_\rho$ range and the resolution range. These parameters are defined in \cref{eq:Presinitial,eq:Densinitial,subsec:list_of_models}. The default resolution of all the weak stratification runs ($H_\rho>1$) is $256^2 \times 384$. The runs with stronger stratification ($H_\rho<1$) or $\mathcal{M}_{\mathrm{bin}}=0.4$ are run with $512^2 \times 768$ resolution elements. We also run four simulations at a resolution of $1024^2\times1536$, for convergence checks and four strongly stratified simulations at $\mathcal{M}=0.01$ and resolution $512^2 \times 768$. Column~4 lists the actual rms $\mathcal{M}$ range, which can be different from the targeted $\mathcal{M}$ in strongly stratified simulations. In column~5, $\mathrm{Fr_\perp}$ refers to the mean perpendicular Froude number of the simulations (see equation~\ref{eq:Fr_perp}). Column~6 shows $\sigma_s^2$, the standard deviation of $s=\ln{\bar{\rho}}$ squared. All quantities ($\mathcal{M}$, $\mathrm{Fr}_\perp$, $\sigma_s$) were averaged over 10 turbulent turnover times, for $6 \leq t/t_\mathrm{turb} \leq 16$. A more detailed version of this table, which lists each of the total of 96 simulations used here, is available in the appendix \cref{tab:sim_list}.\end{footnotesize} 
	
\end{table*}

\subsubsection{Dividing the analysis box into slabs}\label{subsubsec:slab_division}
In order to take into account the variation in $\mathrm{Fr}_\perp$ and $\mathcal{M}$ along $z$ due to steep temperature profiles, and to maintain uniformity during post processing among all our simulations, we divide the the analysis data cube (see definition in \cref{subsubsec:BC}) into four slabs along the $z$ direction ($-0.5\leq z<-0.25$,$-0.25\leq z<0$, $0\leq z<0.25$ and $0.25\leq z\leq0.5$). This means that for each simulation, we have four sets of data-points ($\sigma_s$, $\sigma_{\ln(\bar{P})}$, $\mathcal{M}$, and $\mathrm{Fr}_\perp$), corresponding to each of these four slabs.  

In the presence of significant turbulent pressure ($\propto \text{local }\mathcal{M}^2$), comparable to the thermal pressure, our Dirichlet+reflective boundary conditions break hydrostatic equilibrium. This leads to fluctuations in the computational domain whose amplitude increases for steeper pressure profiles, since the thermal pressure is small at the $z=0.75$ boundary. 
These fluctuations originate at the upper boundary, but they are confined close to the boundary itself, since strong stratification prevents them from travelling to lower $z$. Therefore, as a precaution, we ignore the upper two slabs and only use the lower two slabs in our analysis of simulations that have steep density or pressure profiles ($H_P$, $H_\rho<0.25$).

All our simulations run for a total of 16 eddy turnover times ($t_{\text{eddy}}\approx t_{\text{turb}}$) on the driving length scale. The simulations reach a steady state between $3$--$6\,t_{\text{turb}}$. We analyse turbulence from $6\,t_{\text{turb}}$ to $16\,t_{\text{turb}}$, for a total duration of $10\,t_{\text{turb}}$, for statistical averaging.

% === Results ===
\section{Results and discussion}\label{sec:results-discussion}
Now we describe the results of our simulations and discuss their possible interpretations. We also compare our results against $\delta\bar{\rho}$--$\delta\bar{P}$--$\mathcal{M}$ relations in other stratified turbulence simulations of the ICM.

\subsection{Density and velocity projection maps}\label{subsec:Dens_vel_proj}
\begin{figure*}
		\centering
	\includegraphics[width=1.7\columnwidth]{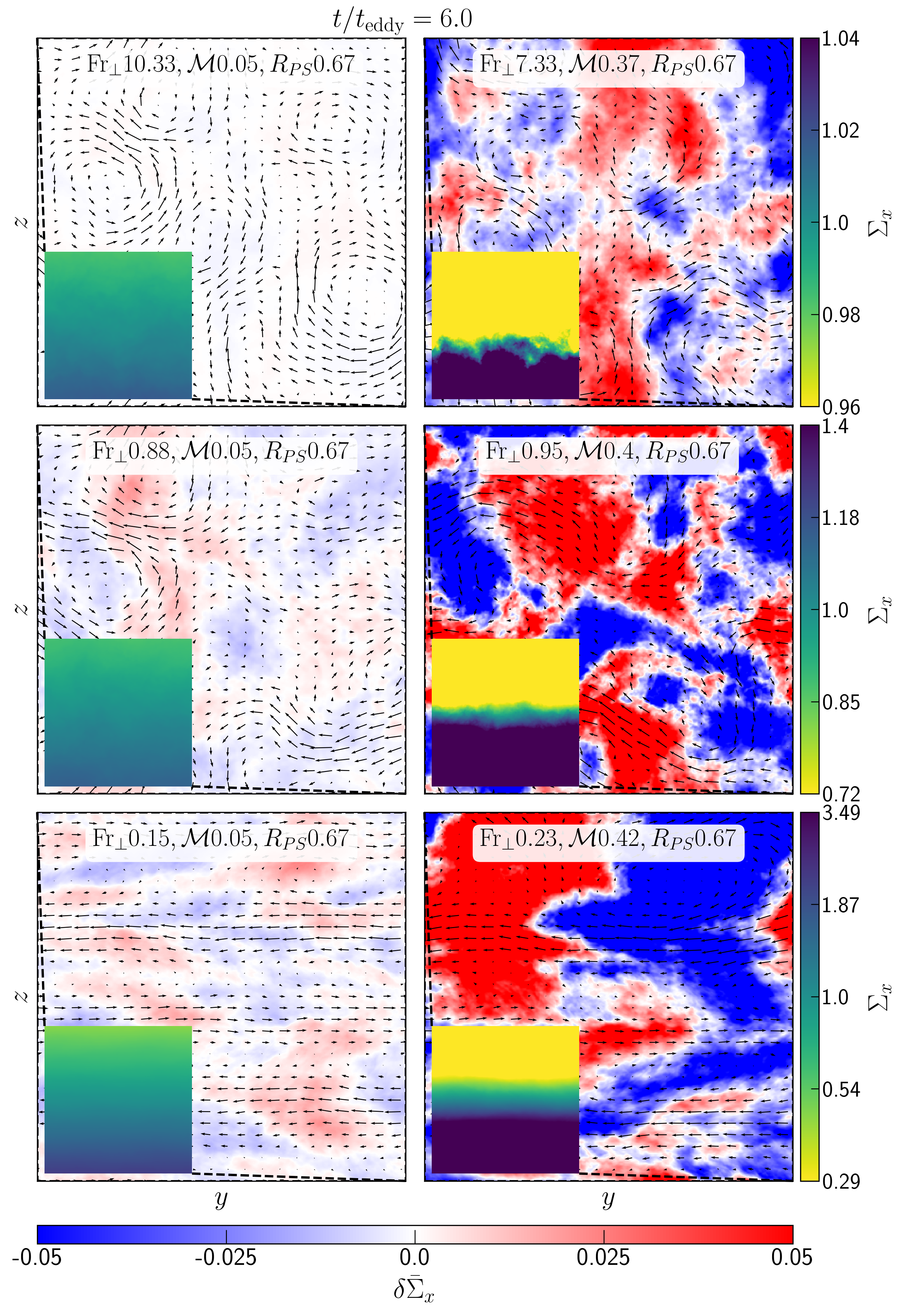}	
	\caption[density-projection plots]{Normalised projected density fluctuations integrated over entire $x-$ extent, $\delta \bar{\Sigma}_x$, with the $x-$integrated velocity field superimposed, for six representative simulations at $t=6\,t_{\mathrm{eddy}}$. The insets show the projected profiles of density, $\Sigma_x$ for the entire box, instead of the fluctuations. All of the 6~simulations shown have the same $R_{PS}=0.67$ but different $H_P$, $H_\rho$ to have roughly the same Fr$_\perp$ in each row and the same $\mathcal{M}$ in each column ($\approx 0.05, 0.4$). The Froude number in each row decreases from $\sim10$ (top panels), to $\sim1$ (middle panels), to $\sim0.2$ (bottom panels). The colourbar for the $\delta \bar{\Sigma}_x$ panels (shown below the sub-panels) has a linear scale. Note that each row of the $\Sigma_x$ insets has its own colourbar, which is in log scale. A movie of the time evolution of these representative simulations is available at this \hyperlink{https://youtu.be/arvouYKQ5Ic}{youtube link}.
	}
	\label{fig:dens-proj-2d}
\end{figure*}

In \cref{fig:dens-proj-2d}, we compare 6 representative simulation models (3 different models with high to low Froude number, from top to bottom, and 2 different Mach numbers, left versus right). Each panel shows the column density fluctuations $\delta\bar{\Sigma}_i=\int\bar{\rho}\mathrm{d}i-1$, with the projected velocity field superimposed as vectors, where $i$ denotes the line of sight (LOS) direction (the absolute column density $\Sigma_i=\int\rho\mathrm{d}i$ is shown in the insets to provide a sense of the strength of the stratification). We have chosen $x$ as the LOS, which is perpendicular to the direction of stratification. For small density fluctuations ($\delta\bar{\rho}<1$), the $\delta\bar{\Sigma}_x$ plots provide a sense of comparison (refer to section~3.2 of \citetalias{Mohapatra2020}) to X-ray surface brightness fluctuations in \cite{zhuravleva2014turbulent}, which have been used to reconstruct turbulent velocities of ICM gas.

For this figure, we have chosen 6 representative simulations, with two different $\mathcal{M}$ and three different $\mathrm{Fr}_\perp$, such that we roughly present the extremes of these two parameters. We have plotted them such that $\mathcal{M}$ is approximately constant along a column, $\mathrm{Fr}_\perp$ is approximately constant along a row and $R_{PS}$ is a constant for all.

For the $\Sigma_x$ plots shown in the insets, we have used a separate log-scale colourbar for each row. The steepness of the $\Sigma_x$ profile is inversely proportional to $H_\rho$. For these plots, $\mathrm{Fr}_\perp\propto\mathcal{M}H_\rho$, as $H_\rho$ decreases between the left and right panels, and also decreases from the top to the bottom panels.  The column density fluctuations ($\delta\bar{\Sigma}_x$) increase with both $\mathcal{M}$ and $\mathrm{Fr}_\perp$, but are  more sensitive to the changes in $\mathcal{M}$. 

In the top and middle rows (weakly and moderately stratified turbulence), we observe that the eddies are roughly circular and hence the velocity field is roughly isotropic (similar to figure~3 of \citetalias{Mohapatra2020}). However, for strongly stratified turbulence shown in the bottom row, the eddies become flatter in the $z$ direction and turbulence forms layered stratified structures. We also observe the correlation between $\delta\bar{\Sigma}_x$ and $v_z$ in the middle row, where upward velocity arrows ($v_z>0$) are associated with regions where $\delta\bar{\Sigma}_x>0$ (shown in red), and downward velocity arrows ($v_z<0$) are associated with regions where $\delta\bar{\Sigma}_x<0$ (shown in blue). This positive correlation implies that some of the kinetic energy is converted into buoyancy potential energy. 

\subsection{Velocity anisotropy}\label{subsec:vel_anisotropy}
\begin{figure}
		\centering
	\includegraphics[width=0.99\columnwidth]{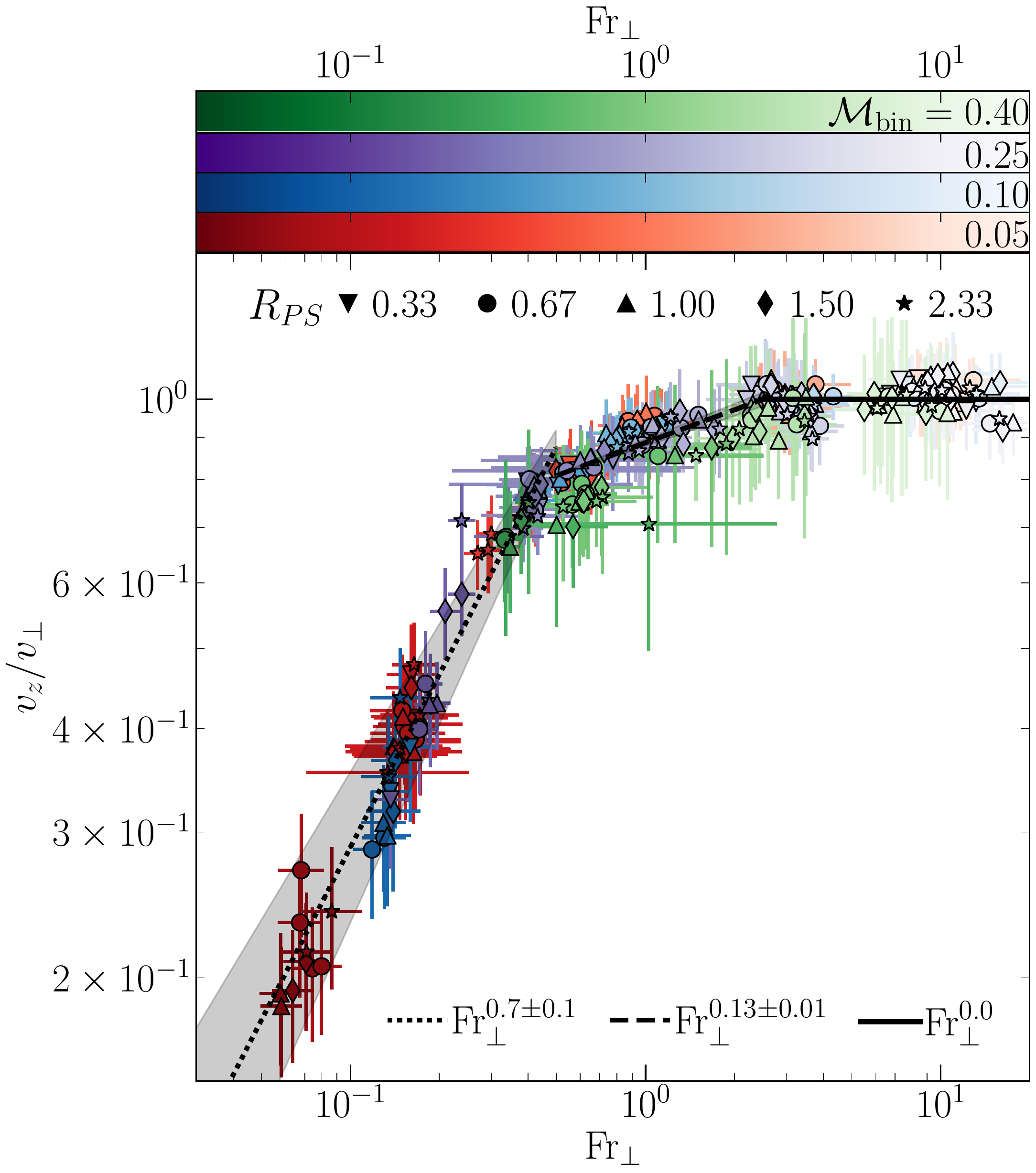}	
	\caption[velratio-Fr plots]{The ratio between $v_z$ and $v_\perp$  versus $\mathrm{Fr}_{\perp}$, where $v_z$ and $v_\perp$ are the rms values of one-component velocities perpendicular and parallel to the direction of stratification, respectively, averaged over slabs in the $z$ direction (see subsubsection~\ref{subsubsec:slab_division}). Colours red, blue, purple and green indicate the $\mathcal{M}_{\mathrm{bin}}$ of the runs. Different shades of these colours indicate the value of $\mathrm{Fr}_{\perp}$, as shown in the colourbars. The different symbols indicate different values of the ratio $R_{PS}=H_P/H_S$. The dotted, dashed, and solid lines indicate fits in the low to high $\mathrm{Fr_\perp}$ regime. The shaded region around the fits shows the $1\sigma$ error margins in the fitting parameters. We see significant anisotropies arising in the turbulence, for strong stratification ($\mathrm{Fr}_{\perp}\lesssim0.5$).}
	\label{fig:velsqratio-Fr}
\end{figure}

In homogeneous idealised turbulence, the velocity field is expected to be isotropic and to follow a nearly Gaussian distribution. However, external fields such as gravity (lower panel of fig.~4 in \citetalias{Mohapatra2020}; \citetalias{Shi2019}) and magnetic fields \citep{Federrath2016JPlPh,Beattie2020MNRAS} can induce strong anisotropies in velocity fields. In our strongly stratified simulations, we expect the $z$ component of velocity $v_z$ to be the most affected by the strength of the stratification.

Understanding the variation in $v_z$ as a function of stratification is of key significance to our study, since in \citetalias{Mohapatra2020} we showed that the additional density fluctuations introduced by buoyancy effects ($\delta\bar{\rho}_\text{buoy}$) are correlated to $v_z$. We also derived a scaling relation for $\delta\bar{\rho}_\text{buoy}$, assuming the velocity field to be isotropic, i.e., $\mean{v_z^2}\approx v^2/3$. However, this assumption breaks down for strongly stratified turbulence ($\mathrm{Fr}_\perp\ll1$), since the velocity field becomes strongly anisotropic and motion is confined to layers perpendicular to the direction of gravity, as we see in the lower panels of \cref{fig:dens-proj-2d}. 

In order to investigate this further and understand the scaling of $v_z$ with $\mathrm{Fr}_\perp$, we show the ratio of $\mean{v_z}_{\mathrm{rms}}$ to $v_\perp$ (defined in \cref{eq:v_perp}) in \cref{fig:velsqratio-Fr}. For weak stratification ($\mathrm{Fr}_\perp\gg1$), the velocity is roughly isotropic ($v_z/v_\perp\approx1$)\footnote{From here on, we denote $\mean{v_z}_{\mathrm{rms}}$ as $v_z$.}, perpendicular and parallel velocities being roughly the same. As we move from right to left in \cref{fig:velsqratio-Fr}, in the moderately stratified turbulence regime ($\mathrm{Fr}_\perp\approx 1$), the ratio starts decreasing slowly with decreasing $\mathrm{Fr}_\perp$. This is expected, as more of the $z$ direction kinetic energy gets converted into buoyancy potential energy with increasing strength of stratification. As we move further left in the plot to the strong stratification limit ($\mathrm{Fr}_\perp\ll1$), the ratio shows a sharp decrease with decreasing $\mathrm{Fr}_\perp$, with $v_z/v_\perp\propto\mathrm{Fr}_\perp^{0.7}$. 

For strongly stratified turbulence, buoyancy dominates on large scales, for $\ell_O<\Tilde{\ell}<L_{\text{driv}}$ (defined in eq.~\ref{eq:l_Ozmidov}), till the relative strengths of buoyancy and turbulence terms become equal. For $\Tilde{\ell}<\ell_O$, the turbulence transitions to weakly stratified turbulence. Thus, unlike isotropic turbulence, $v_z$ is set by $\Tilde{v}_\parallel(\ell_O)$ instead of the integral scale velocity $v_\parallel$. For $\Tilde{\ell}>\ell_O$, we may use the Boussinesq approximation (changes in density are small relative to the mean density), for which \cref{eq:Euler1} reduces to $\nabla\cdot\mathbf{v}=0$. Then the ratio
\begin{equation}
    \frac{\Tilde{v}_\parallel\left(\Tilde{\ell}_\perp\right)}{{\Tilde{v}_\perp\left(\Tilde{\ell}_\perp\right)}}\approx \frac{\Tilde{\ell}_\parallel}{\Tilde{\ell}_\perp}\approx\frac{\Tilde{\mathrm{Fr}}_\perp\left(\Tilde{\ell}_\perp\right)}{\Tilde{\mathrm{Fr}}_\parallel\left(\Tilde{\ell}_\perp\right)} \approx\Tilde{\mathrm{Fr}}_\perp(\Tilde{\ell}_\perp) = \frac{\Tilde{v}_\perp\left(\Tilde{\ell}_\perp\right)}{N\Tilde{\ell}_\perp},\label{eq:v_scale_ratio}
\end{equation}
using \cref{eq:scale_Fr_perp,eq:scale_Fr_parallel}, where we assume $\tilde{\rm Fr}_\parallel \sim 1$ for strong stratification and that the vertical velocity fluctuations peak at the Ozmidov scale and not the driving scale. This gives us 
\begin{equation}
    \frac{v_z}{v_\perp}\approx\frac{\Tilde{v}_\parallel\left(\ell_O\right)}{v_\perp}\approx \frac{\Tilde{v}_\perp^2(\ell_O)}{N\ell_O v_\perp} =\frac{\epsilon_K^{2/3}}{\ell_O^{1/3}Nv_\perp}=\mathrm{Fr}_\perp^{0.5},\label{eq:vz_scaling}
\end{equation}
using \cref{eq:l_Ozmidov,eq:v_scale_ratio} \citep[see example 14.2 in][]{davidson_2013}. We observe $v_z/v_\perp\propto\mathrm{Fr}_\perp^{0.7\pm0.1}$ in our strongly stratified simulations, which roughly agrees with the theoretical prediction. Some deviations from the theoretical prediction may arise because the low-Froude simulations are not in the limit of $\mathrm{Fr}_\perp \ll1$, but around $\mathrm{Fr}_\perp\sim0.1$.

\subsection{Density and pressure fluctuations}\label{subsec:dens_fluc}
In this subsection, we discuss the variation of density and pressure fluctuations as a function of our simulation parameters - $\mathcal{M}$, $\mathrm{Fr}_\perp$ and $R_{PS}$. The scaling relations between these quantities are important for obtaining ICM gas velocity estimates from X-ray and SZ observations. The X-ray surface brightness fluctuations and SZ effect fluctuations are used to calculate the amplitude of density and pressure fluctuations, respectively, which are further used to calculate velocity fluctuations using these relations \citep{zhuravleva2013quantifying,zhuravleva2014relation,khatri2016}. 

\subsubsection{Density and pressure fluctuations as a function of $\mathcal{M}$}\label{subsubsec:dens_pres_fluc_Mach}
\begin{figure}
		\centering
	\includegraphics[width=0.99\columnwidth]{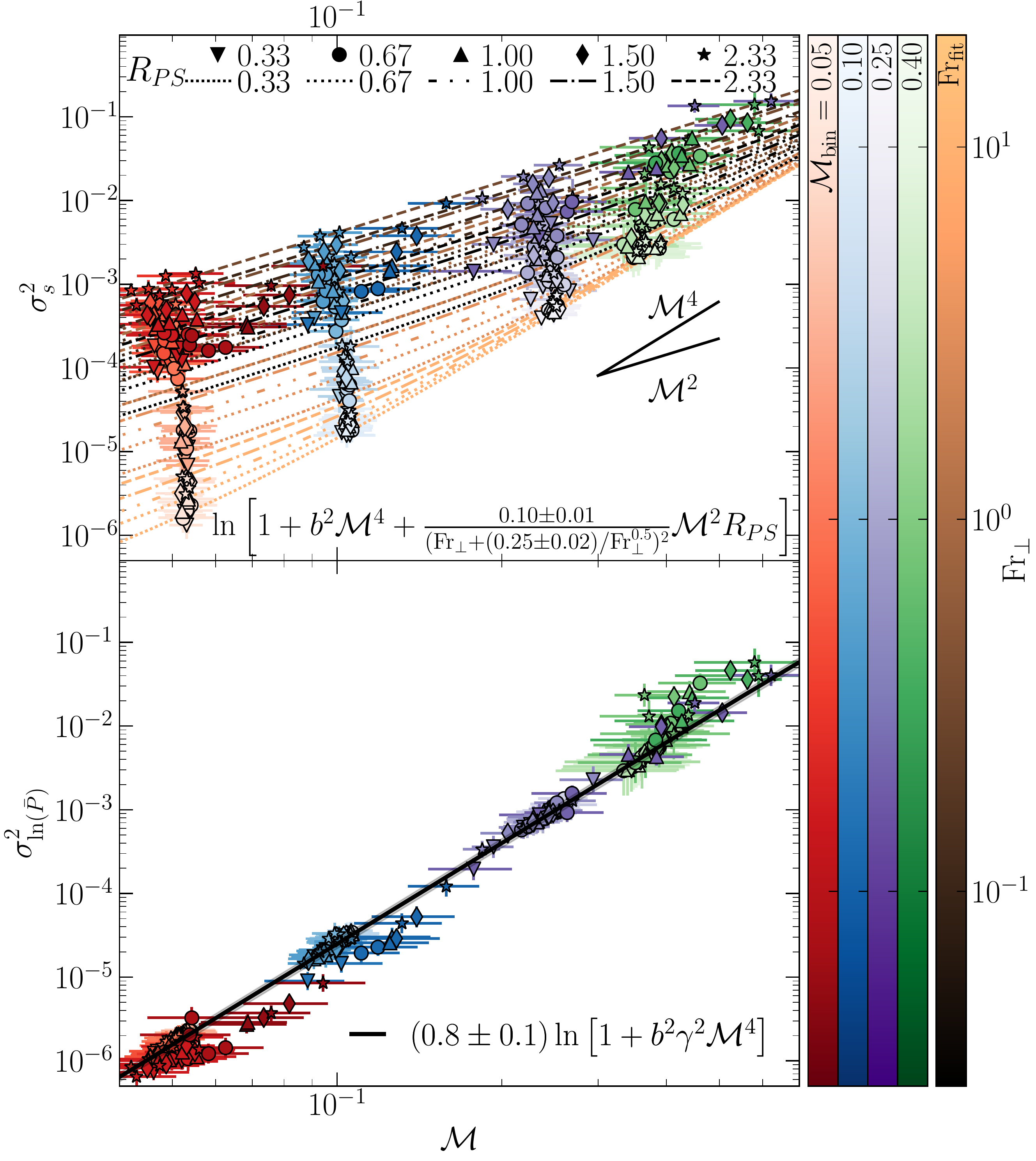}	
	\caption[sig-mach plots]{ The density and pressure fluctuations ($\sigma_{\mathrm{s}}^2$; upper panel and $\sigma_{\log{\bar{P}}}^2$; lower panel) versus the rms Mach number ($\mathcal{M}$) for all our simulations. The rightmost colourbar indicates the value of $\mathrm{Fr}_{\perp}$ used for plotting the different fits in the upper panel. The different line styles indicate the value of the ratio $R_{PS}=H_P/H_S$ used for the different fits in the upper panel. The shades on different symbols correspond to the different Froude numbers. Density fluctuations are smaller for a larger Fr$_\perp$ (weaker stratification). The respective fitting functions (\cref{subsubsec:dens_pres_fluc_Froude}) for the upper and lower panels are indicated in the bottom right corner of each panel. For reference, we have also shown sample $\mathcal{M}^2$ and $\mathcal{M}^4$ scaling in the upper panel.}
	\label{fig:sig-mach}
\end{figure}

In \cref{fig:sig-mach}, we show the density and pressure fluctuations squared ($\sigma_s^2$ and $\sigma_{\ln{\bar{P}}}^2$) versus $\mathcal{M}$ for all our simulations. We also present approximate fits based on the scaling relations that we propose in the next section \ref{subsubsec:dens_pres_fluc_Froude}. Clearly, $\sigma_s^2$ increases with increasing $\mathcal{M}$, stratification strength (decreasing $\mathrm{Fr}_\perp$) and $R_{PS}$. In comparison, $\sigma_{\ln{\bar{P}}}^2$ only increases with $\mathcal{M}$ and is independent of $\mathrm{Fr}_\perp$ and $R_{PS}$. 

As we discussed in section~3.5 of \citetalias{Mohapatra2020}, the density fluctuations are comprised of two components, which can be written as the sum of an un-stratified and a stratified turbulence component:
\begin{equation}
    \delta\bar{\rho}^2=\delta\bar{\rho}_{\text{turb}}^2+\delta\bar{\rho}_\text{buoy}^2 \label{eq:del_rho_components},
\end{equation}
where $\delta\bar{\rho}_{\text{turb}}^2$ scales as $b^2\mathcal{M}^4$ (\citealt{Mohapatra2019}; \citetalias{Mohapatra2020}), and $b$ is the turbulence driving parameter \citep{Federrath2008}. The parameter $b=1/3$ for solenoidal turbulence, as we use for the present set of simulations. Our fits in \cref{fig:sig-mach} show this dependence on $\mathcal{M}$ for $\mathrm{Fr}_{\text{fit}}\gg1$ (light bronze coloured fits). We also showed that $\delta\bar{\rho}_{\text{turb}}$ corresponds to adiabatic density fluctuations, so the corresponding pressure fluctuations $\delta{\bar{P}}^2$ scale as $\gamma^2\delta\bar{\rho}^2_{\text{turb}}\propto\mathcal{M}^4$ for $\mathrm{Fr}_\perp\gg1$. 
The $\delta\bar{\rho}^2_{\text{turb}}$ component of density fluctuations dominates for weakly stratified turbulence or at large $\mathcal{M}$. 

For moderate stratification ($\mathrm{Fr}_\perp\sim1$) or low $\mathcal{M}$ turbulence, the $\delta\bar{\rho}_\text{buoy}$ term dominates, but it corresponds to the isobaric motions of isotropic gas parcels, which have zero contribution to the net pressure fluctuations. Hence $\delta\bar{P}$ still scales as $\gamma\delta\bar{\rho}_{\text{turb}}$ and shows a $\mathcal{M}^4$ variation throughout as seen in the lower panel of \cref{fig:sig-mach}. This scaling seems to hold even in the strongly stratified turbulence limit, which means that the nature of $\delta\bar{\rho}_{\text{buoy}}$ is isobaric even for $\mathrm{Fr}\ll1$. This behaviour is expected, since a rising/falling parcel of gas with subsonic velocity is always in pressure equilibrium with its immediate surroundings (such that $\delta \bar{P}_{\rm buoy}=0$). Thus, the expression for logarithmic pressure fluctuations becomes
\begin{equation}
    \sigma_{\ln\bar{P}}^2=\ln\left[1+b^2\gamma^2\mathcal{M}^4\right].\label{eq:siglnp-mach}
\end{equation}
In the lower panel of \cref{fig:sig-mach}, we fit the data to \cref{eq:siglnp-mach} using the fitting tool LMfit \citep{LMfit2016ascl.soft06014N}. The results are in good agreement with our expectations.

In \citetalias{Mohapatra2020}, we also showed that $\delta\bar{\rho}_\text{buoy}^2$ increases with $\mathcal{M}$ as approximately $\mathcal{M}^2$, for constant $\mathrm{Fr}_\perp\gtrsim1$ (or $\mathrm{Ri}\lesssim1$). The motions in the $z$ direction associated with $\delta\bar{\rho}_\text{buoy}$ are strongly constrained for $\mathrm{Fr}_\perp\ll1$ and because of the large energy cost, BV oscillations (with small displacement in the $z$ direction) dominate over turbulence in the vertical direction. But the scaling with $\mathcal{M}$ still holds in this limit, as is seen in the dark bronze fits in the upper panel of \cref{fig:sig-mach}. 

\subsubsection{Density \& pressure fluctuations as a function of $\mathrm{Fr}_\perp$ and $R_{PS}$}\label{subsubsec:dens_pres_fluc_Froude}
\begin{figure}
		\centering
	\includegraphics[width=0.99\columnwidth]{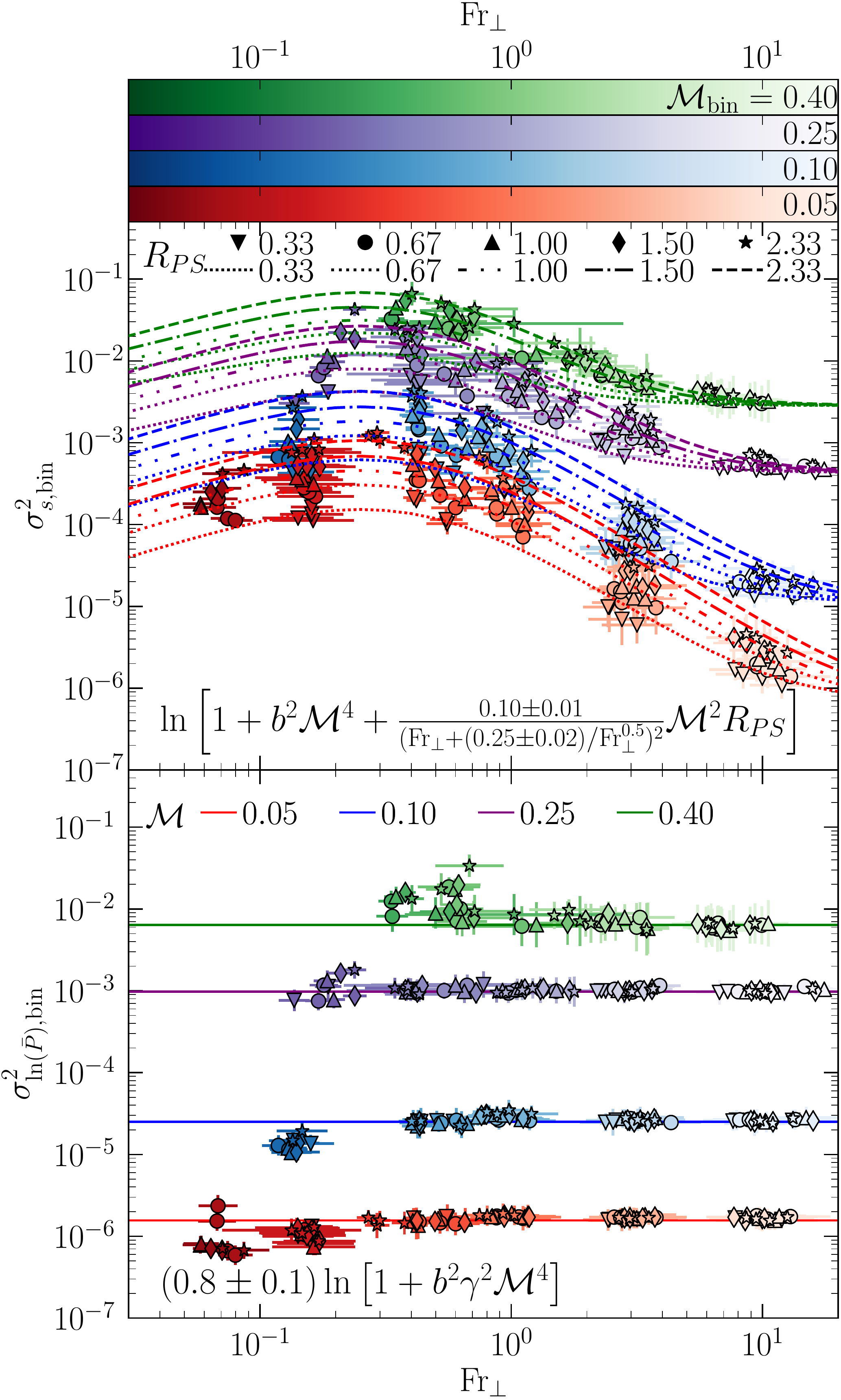}	
	\caption[sigcorr-Fr plots]{ The density and pressure fluctuations ($\sigma_{s\mathrm{,bin}}^2$; upper panel and $\sigma_{\log{\bar{P}}\mathrm{,bin}}^2$; lower panel) versus the transverse Froude number ($\mathrm{Fr}_{\perp}$) for all our simulations. The symbol shape corresponds to $R_{PS}$ and the shade indicates the Mach number. The different line styles indicate the value of the ratio $R_{PS}$ used for our different fits in the upper panel. The fits are given in individual panels.}
	\label{fig:sigcorr-Fr}
\end{figure}

In this subsection, we discuss the scaling of density and pressure fluctuations with the stratification parameters $\mathrm{Fr}_\perp$ and $R_{PS}$. We know that the $\delta\bar{\rho}_{\text{turb}}^2$ and $\delta\bar{\rho}_{\text{buoy}}^2$ terms scale as $\mathcal{M}^4$ and $\mathcal{M}^2$, respectively. In order to make comparisons between different $\mathrm{Fr}_\perp$ and $R_{PS}$ easier, we normalise $\sigma_s^2$ and $\sigma_{\ln\bar{P}}^2$ to $\sigma_{s,\text{bin}}^2$ and $\sigma_{\ln \bar{P},\text{bin}}^2$, respectively, given by
\begin{subequations}
\begin{align}
    &\sigma_{s,\text{bin}}^2=\ln\left[1+b^2\mathcal{M}_{\text{bin}}^4+\frac{\mathcal{M}_{\text{bin}}^2}{\mathcal{M}^2}\left(\exp{\sigma_s^2}-1-b^2\mathcal{M}^4\right)\right]\label{eq:sigs_bin}, \\
    &\sigma_{\ln\bar{P},\text{bin}}^2 =\ln\left[1+\frac{\mathcal{M}_{\text{bin}}^4}{\mathcal{M}^4}\left(\exp{\sigma_{\ln\bar{P}}^2}-1\right)\right].\label{eq:siglnP_bin}
\end{align}
\end{subequations}
This way, the runs with different $\mathcal{M}$ are scaled to the same $\mathcal{M}_{\text{bin}}$. We described $\mathcal{M}_{\text{bin}}$ in \cref{subsec:list_of_models} and $\mathcal{M}_{\text{bin}}\approx\mathcal{M}$ for shallow initial density and pressure profiles ($H_P,H_\rho\gtrsim1$). We take four different values of $\mathcal{M}_{\text{bin}}$, namely $\mathcal{M}_{\text{bin}}=0.05$, $0.10$, $0.25$, and $0.40$. We show $\sigma_{s,\text{bin}}^2$ and $\sigma_{\ln\bar{P},\text{bin}}^2$ vs $\mathrm{Fr}_\perp$ in \cref{fig:sigcorr-Fr}.  

As we move from right to left in this plot (increasing stratification and decreasing $\mathrm{Fr}_\perp$), $\sigma_{s,\text{bin}}^2$ increases for $\mathrm{Fr}_\perp\gtrsim0.5$. It reaches a peak around $\mathrm{Fr}_\perp\approx0.5$ and starts decreasing for smaller $\mathrm{Fr}_\perp$. For $\mathrm{Fr}_\perp\lesssim1$, $\sigma_{s,\text{bin}}^2$ increases with $R_{PS}$, but this dependence is weaker as compared to that on $\mathrm{Fr}_\perp$. For a constant $\mathcal{M}_{\text{bin}}$, $\sigma_{\ln\bar{P},\text{bin}}^2$ stays constant with both $\mathrm{Fr}_\perp$ and $R_{PS}$, as expected from \cref{eq:siglnp-mach}.

All the variation in $\sigma_{s,\text{bin}}^2$ can be attributed to the dependence of $\delta\bar{\rho}_{\text{buoy}}^2$ on these stratification parameters, since $\delta\bar{\rho}_{\text{turb}}^2$ depends only on $\mathcal{M}$. The $\delta\bar{\rho}_{\text{buoy}}^2$ component is correlated with the rms displacement in the $z$ direction $\mean{\delta z^2}$ and is given by
\begin{subequations}
\begin{equation}
    \delta\bar{\rho}_{\text{buoy}}^2=\frac{N^4}{g^2}\mean{\delta z^2} \label{eq:delrho_buoy}.
\end{equation}
In the weak and moderately stratified turbulence limit ($\mathrm{Fr}_\perp\gtrsim1$), we simplify this expression further and showed in \citetalias{Mohapatra2020} that
\begin{equation}
    \delta\bar{\rho}_{\text{buoy}}^2\approx\zeta^2\mathcal{M}^2\mathrm{Ri}R_{PS}\approx\zeta^2\frac{\mathcal{M}^2R_{PS}}{\mathrm{Fr}_\perp^2},  \label{eq:delrho_buoy-weak}
\end{equation}
where $\zeta^2$ is a fitting parameter.
However, this simplification involved assuming the velocity field to be roughly isotropic, which is true for $\mathrm{Fr}_\perp\gtrsim 0.5$ (see fig.~\ref{fig:velsqratio-Fr}), not for strongly stratified turbulence. In the limit of strongly stratified turbulence ($\mathrm{Fr}_\perp\ll1$), the $z$ direction motions are heavily suppressed by buoyancy and the velocity field is no longer isotropic, as we showed in \cref{subsec:vel_anisotropy}. In this limit, 
\begin{align}
    \delta\bar{\rho}_{\text{buoy}}^2 &\approx\frac{N^4}{g^2}\mean{\delta z^2}\approx\frac{N^4}{g^2}\zeta^{\prime 2}\frac{v_z^2}{N^2}\nonumber\\
    &\approx \zeta^{\prime 2}\mathcal{M}^2\mathrm{Fr}_\perp R_{PS},\label{eq:delrho_buoy-strong}
\end{align}
using \cref{eq:grav,eq:BVoscilations,eq:vz_scaling}. Interpolating between the two asymptotic  expressions for $\delta\bar{\rho}_{\text{buoy}}$ in \cref{eq:delrho_buoy-weak} and \cref{eq:delrho_buoy-strong}, we get
\begin{equation}
    \delta\bar{\rho}_{\text{buoy}}^2=\frac{\zeta_1^2\mathcal{M}^2R_{PS}}{\left(\mathrm{Fr}_\perp+\zeta_2/\sqrt{\mathrm{Fr}_\perp}\right)^2}\label{eq:delrho_buoy_combined},
\end{equation}
where $\zeta_1$ and $\zeta_2$ are fitting parameters. The combined expression for the net density fluctuations is then given by
\begin{align}
    &\delta\bar{\rho}^2=b^2\mathcal{M}^4+\frac{\zeta_1^2\mathcal{M}^2R_{PS}}{\left(\mathrm{Fr}_\perp+\zeta_2/\sqrt{\mathrm{Fr}_\perp}\right)^2}\text{ and} \label{eq:delrho_combined}\\
    &\sigma_s^2=\ln\left[1+b^2\mathcal{M}^4+\frac{\zeta_1^2\mathcal{M}^2R_{PS}}{\left(\mathrm{Fr}_\perp+\zeta_2/\sqrt{\mathrm{Fr}_\perp}\right)^2}\right].\label{eq:sigs_combined}
\end{align}
\end{subequations}
Here we substituted $\sigma_s^2=\ln\left[1+\sigma_{\bar{\rho}}^2\right]$. This is valid for log-normal distributions of $\bar{\rho}$ but it also holds approximately for non-log-normal distributions with small $\sigma_{\bar{\rho}}^2$ (see Appendix~\ref{app:sigma_rho}).

Thus, we find that introducing stratification has two main effects. For $\mathrm{Fr}_\perp\gtrsim0.5$, due to the existing density gradient, the turbulent motions in the $z$ direction produce higher density fluctuations. This happens simply because when a gas parcel moves along the $z$ direction, it only attains pressure equilibrium at that height, but still has a density contrast with respect to its surroundings.
On assuming isotropic gas velocities, which is roughly valid for $\mathrm{Fr}\gtrsim0.5$ (see \cref{fig:velsqratio-Fr}), we obtain $\delta\bar{\rho}_{\mathrm{buoy}}^2=\zeta^2\mathcal{M}^2 R_{PS}/\mathrm{Fr}_\perp^2$ (eq.~\ref{eq:delrho_buoy-weak}). However, on further increasing the stratification ($\mathrm{Fr}_\perp\lesssim0.5$), the turbulence becomes anisotropic as gas motions along the $z$ direction are suppressed, with the kinetic energy along the $z$ direction being converted into buoyancy potential energy. The turbulent eddies flatten along the $z$ direction and become pancake-like (see lower panels of \cref{fig:dens-proj-2d}). In this limit, the motion along the $z$ direction is best described by BV oscillations. The amplitude of these oscillations is proportional to $v_z$, which decreases with decreasing $\mathrm{Fr}_\perp$ for constant $\mathcal{M}$ (see fig. \ref{fig:vz-vperp-Fr-mach001}). Substituting the $\mathrm{Fr}_\perp$ dependence of the ratio $v_z/v_\perp$, we obtain $\delta\bar{\rho}_{\mathrm{buoy}}^2=\zeta^{\prime2}\mathcal{M}^2 R_{PS}\mathrm{Fr}_\perp$ (eq.~\ref{eq:delrho_buoy-strong}). The general expression for the dependence of $\delta\bar{\rho}_{\mathrm{buoy}}^2$ (eq.~\ref{eq:delrho_buoy_combined}) is thus an interpolation between these two forms.

For even stronger stratification ($\mathrm{Fr}_\perp\lesssim0.001$), we expect the amplitude of $\delta\bar{\rho}_{\mathrm{buoy}}^2$ to decrease below $\delta\bar{\rho}_{\mathrm{turb}}^2$, and the net density fluctuations would again be given by $\delta\bar{\rho}^2\sim b^2\mathcal{M}^4$, similar to the unstratified subsonic turbulence scaling. One can interpret this limit as 2D subsonic turbulence, with $\mathcal{M}$ representing the 2D Mach number, since $v_z\ll v_\perp$.

\subsubsection{Obtaining the fitting parameters}\label{subsubsec:fitting_parameters}

\begin{figure}
		\centering
	\includegraphics[width=0.99\columnwidth]{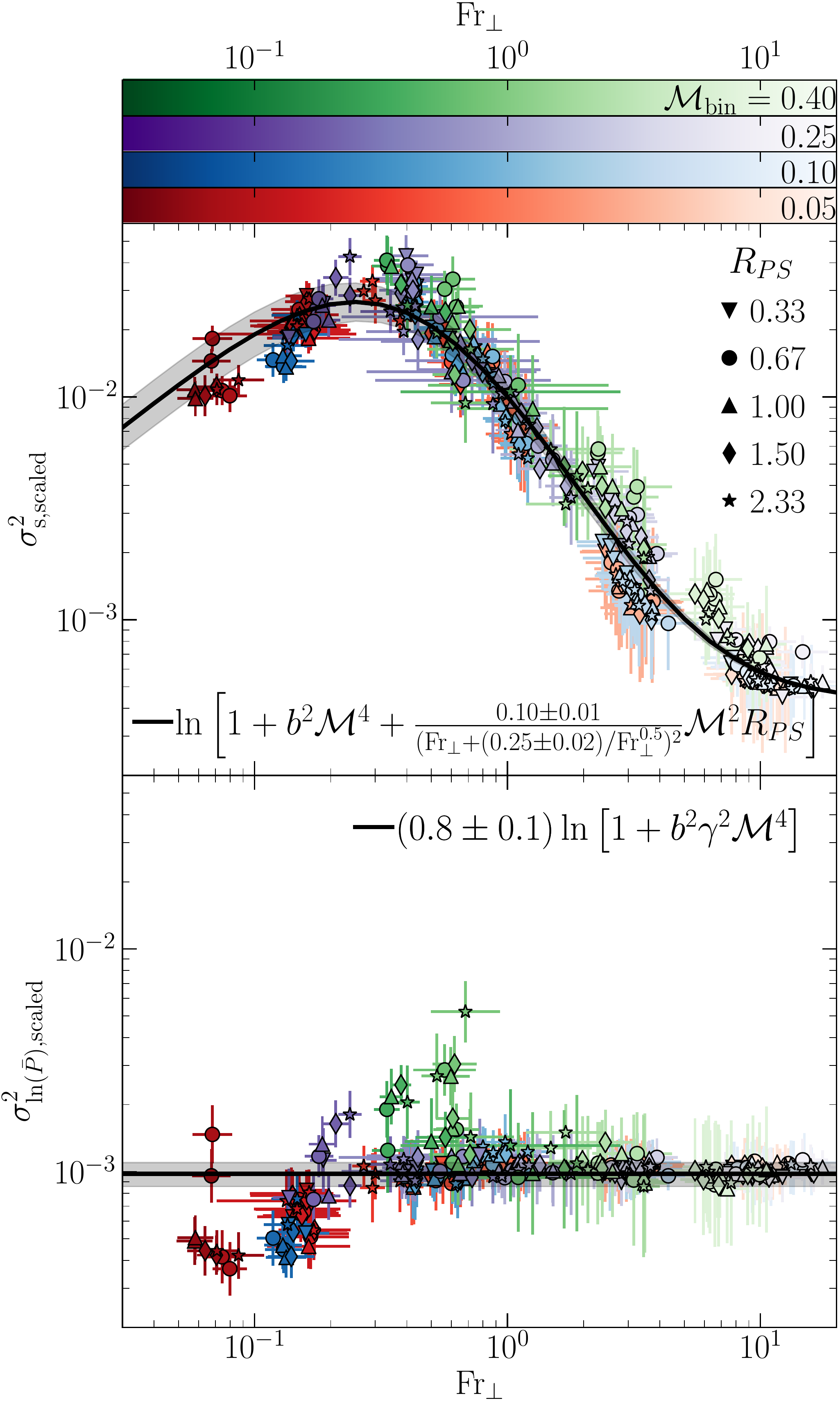}	
	\caption[sig-oneMach-Fr plots]{The density ($\sigma_{s\mathrm{,scaled}}^2$; upper panel) and pressure fluctuations ($\sigma_{\log{\bar{P}\mathrm{,scaled}}}^2$; lower panel), scaled such that they collapse on to a single line as a function of Fr$_\perp$ all our simulations. The symbol shape corresponds to $R_{PS}$ and the colours indicate the Mach number. The fits are shown in the respective panels.  The shaded region around the fits shows the $1\sigma$ error margins in the fitting parameters.
	}
	\label{fig:sig-oneMach-Fr}
\end{figure}
Here we derive the fitting parameters $\zeta_1$ and $\zeta_2$, such that they describe the variation in density fluctuations for all of our 96 simulations simultaneously. We use the fitting tool LMfit with \cref{eq:sigs_combined} as the fitting function, $\mathcal{M}$, $\mathrm{Fr}_\perp$ and $R_{PS}$ as independent variables, and $\sigma_s^2$ as the dependent variable. We obtain $\zeta_1^2=0.10\pm0.01$ and $\zeta_2=0.25\pm0.02$. 

In order to present all our data and the fitting function together, we scale all density and pressure fluctuations to $\mathcal{M}_{\text{scaled}}=0.25$ and $R_{PS}=2.33$. These are the same parameter values we used in all simulations of \citetalias{Mohapatra2020} and hence they provide a direct comparison between both of our studies. We construct two new scaled quantities, given by
\begin{subequations}
\begin{alignat}{2}
    &\sigma_{s,\text{scaled}}^2 &=\ln\left[1+b^2(0.25)^4+\left(\frac{0.25}{\mathcal{M}}\right)^2\frac{2.33}{R_{PS}}\right.\nonumber\\
    &&\left.\vphantom{\left(\frac{0.25}{\mathcal{M}}\right)^2}\times\left(\exp{\sigma_s^2}-1-b^2\mathcal{M}^4\right)\right]\text{ and}\label{eq:sigs_scaled}\\
    &\sigma_{\ln\bar{P},\text{scaled}}^2 &=\ln\left[1+\left(\frac{0.25}{\mathcal{M}}\right)^4\left(\exp{\sigma_{\ln\bar{P}}^2}-1\right)\right].\label{eq:siglnP_scaled}
\end{alignat}
\end{subequations}
In \cref{fig:sig-oneMach-Fr}, we show these two quantities as a function of $\mathrm{Fr}_\perp$. We also show the fitting function \cref{eq:sigs_combined} for $\mathcal{M}=0.25$ and $R_{PS}=2.33$. Thus, our scaling relation for $\sigma_s^2$ works well for the entire range of parameter space we have scanned using the 96 simulations.
The value of $\sigma_{\ln\bar{P},\text{scaled}}^2$ is roughly constant, as expected from \cref{eq:siglnp-mach}. We expect the variation to be mostly due to two reasons both leading to inaccurate estimates of $\mathcal{M}$, and these variations are amplified in the plot since $\sigma_{\ln\bar{P},\text{scaled}}^2\propto\mathcal{M}^4$. Firstly, the box heats up in larger $\mathcal{M}_{\text{bin}}$ simulations (especially $\mathcal{M}_{\text{bin}}=0.25$ and $0.4$), which leads to an increased speed of sound and a decreased Mach number. Secondly, due to the steep temperature profiles for simulations with $H_P,H_\rho<0.2,$ $(H_P\neq H_\rho)$, the speed of sound significantly varies along the $z$ direction, whereas we drive an isotropic velocity field. This leads to a strong variation in $\mathcal{M}$ along the $z$ direction, which may cause the differences from the scaling relations for $\mathrm{Fr}_\perp\lesssim0.5$. 

\subsubsection{Pressure fluctuations and the SZ effect}\label{subsubsec:PresSZ}
In this subsection we discuss the importance of pressure fluctuations for estimating turbulent gas velocities in the ICM. In the previous subsections, we showed that density fluctuations are sensitive to the parameters $\mathrm{Fr}_\perp$ and $R_{PS}$, whereas pressure fluctuations depend only on $\mathcal{M}$. We found that the relation $\sigma_{\bar{P}}^2=b^2\gamma^2\mathcal{M}^4$ describes all our simulations very well (see lower panels of \cref{fig:sig-mach} and \cref{fig:sig-oneMach-Fr}). This happens because the stratified turbulence component of density fluctuations is isobaric (see figure~8 of \citetalias{Mohapatra2020}). In figure~14 of \cite{Mohapatra2019}, we also showed that pressure fluctuations do not significantly depend on whether radiative cooling is included or not. While density fluctuations are sensitive to the thermodynamic and stratification parameters, pressure fluctuations are independent of these. Thus, compared to density fluctuations, we expect pressure fluctuations to provide a  robust estimate of turbulent velocities.

Hot electrons in the ICM, on average, up-scatter the 
 cosmic microwave background (CMB) photons through inverse Compton scattering, and the change in the CMB brightness temperature is proportional to the ICM electron pressure integrated along the line of sight. This effect is known as the thermal Sunyaev-Zeldovich (tSZ) effect \citep[see ][ for a review]{Mroczkowski2019}. 
 Although it is possible to recover turbulent gas velocities from the resolved observations of the tSZ effect \citep{khatri2016}, the present observations suffer from a lack of angular resolution (e.g., up to a few $100$~$\mathrm{kpc}$ for the Coma cluster).
The tSZ observations also suffer from projection effects (more than X-ray surface brightness which is proportional to $\rho^2$), which makes the measured pressure fluctuations even smoother than the density fluctuations. Many of these limitations are expected to be addressed by future facilities \citep[see section~6.2 of][]{Mroczkowski2019}. The relative robustness of the relation between pressure fluctuations and turbulent velocities (unlike density fluctuations which depend on stratification and thermodynamic parameters) provides a strong motivation for high angular resolution SZ observations.

\subsubsection{Testing the scaling relation at lower $\mathcal{M}$ and $\mathrm{Fr}_\perp\ll 1$}\label{subsubsec:testing_low_Fr}

\begin{table}
	\centering
	\def\arraystretch{1.35}
	\caption{Simulation parameters for the $\mathcal{M}=0.01$, $\mathrm{Fr}_\perp<0.4$ runs}
	\label{tab:lowMach_sim_params}
	\resizebox{\columnwidth}{!}{
		\begin{tabular}{lcccc} 
			\hline%To colour these cyan
			Label & $H_{\rho}$ & $\mathcal{M}$ & $\sigma_s^2$ & $\sigma^2_{\log\bar{P}}$\\
			(1) & (2) & (3) & (4) & (5)\\
			\hline
			$\mathrm{Fr}0.34\mathcal{M}0.01R_{PS}0.67$ & $8.0$ & $(8.7\pm0.3)\times10^{-3}$ & $1.2^{-0.2}_{+0.2}\times10^{-5}$ & $1.4^{-0.3}_{+0.4}\times10^{-9}$\\
			$\mathrm{Fr}0.18\mathcal{M}0.01R_{PS}0.67$ & $4.0$ & $(8.9\pm0.4)\times10^{-3}$ &  $1.1^{-0.1}_{+0.2}\times10^{-5}$ & $1.4^{-0.2}_{+0.3}\times10^{-9}$\\
			$\mathrm{Fr}0.09\mathcal{M}0.01R_{PS}0.67$ & $2.0$ & $(9.9\pm0.8)\times10^{-3}$ &  $7.0^{-1.2}_{+1.5}\times10^{-6}$ & $1.6^{-0.4}_{+0.6}\times10^{-9}$\\
			$\mathrm{Fr}0.04\mathcal{M}0.01R_{PS}0.67$ & $1.0$ & $(1.0\pm0.1)\times10^{-2}$ &  $6.4^{-1.1}_{+1.4}\times10^{-6}$ & $2.5^{-0.6}_{+0.8}\times10^{-9}$\\
		\hline
			
	\end{tabular}}
	\justifying \\ \begin{footnotesize} Notes: All these simulations have grid resolution $512^2\times768$. The columns (2) - (5) have their usual meanings. For these runs, $H_P/H_\rho=1$.\end{footnotesize} 
	
\end{table}

\begin{figure}
		\centering
	\includegraphics[width=0.99\columnwidth]{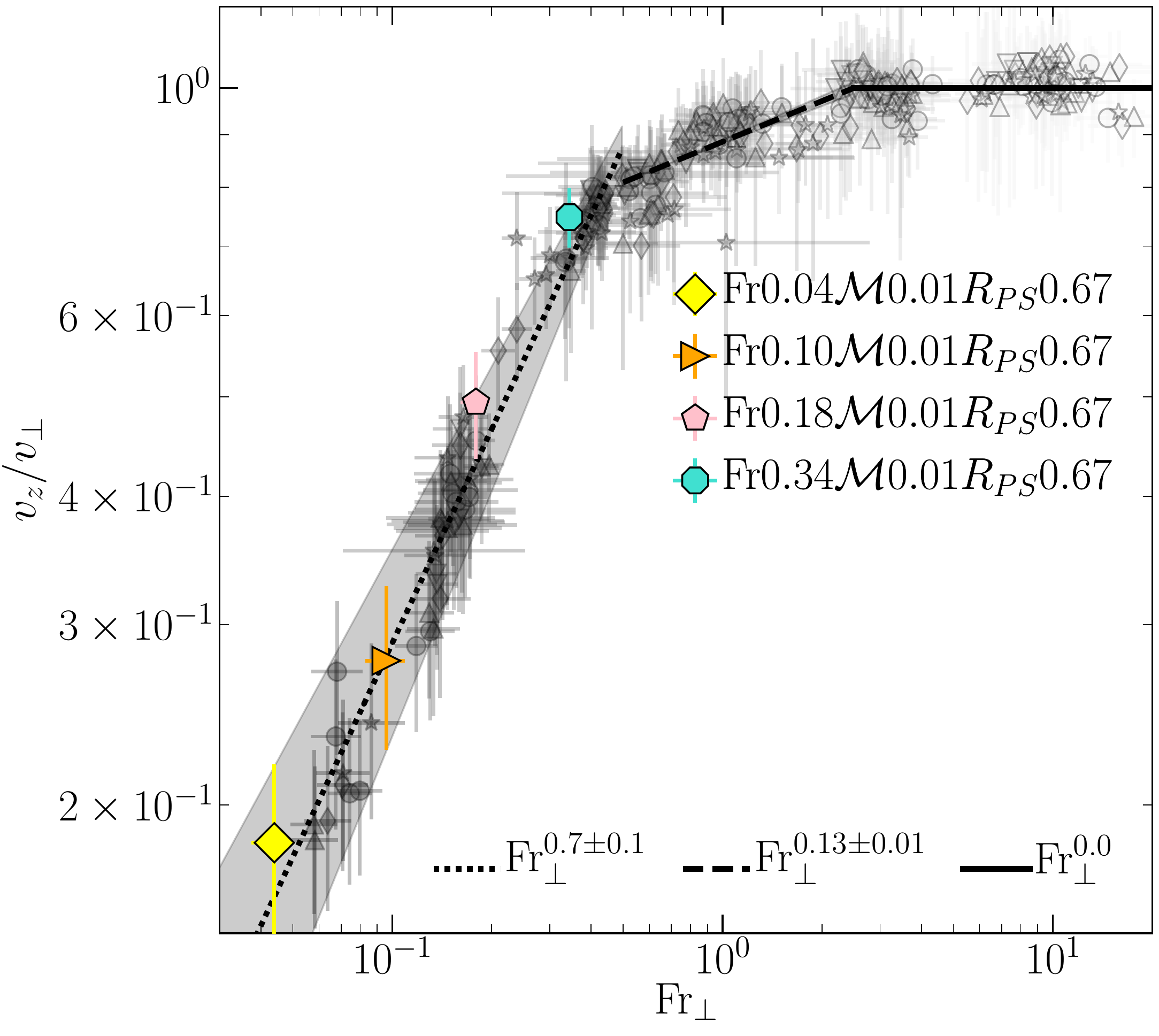}	
	\caption[velz-Fr-relation at low mach]{The ratio between $v_z$ and $v_\perp$  versus $\mathrm{Fr}_{\perp}$ for our simulations with $\mathcal{M}\approx0.01$, $R_{PS}=0.67$, and $\mathrm{Fr}_\perp\lesssim0.4$. We also show our data points from \cref{fig:velsqratio-Fr} in grey. The low-$\mathcal{M}$ data follow the $\mathrm{Fr}_\perp^{0.7}$ scaling for $\mathrm{Fr}_\perp\lesssim0.4$.}
	\label{fig:vz-vperp-Fr-mach001}
\end{figure}

\begin{figure}
		\centering
	\includegraphics[width=0.99\columnwidth]{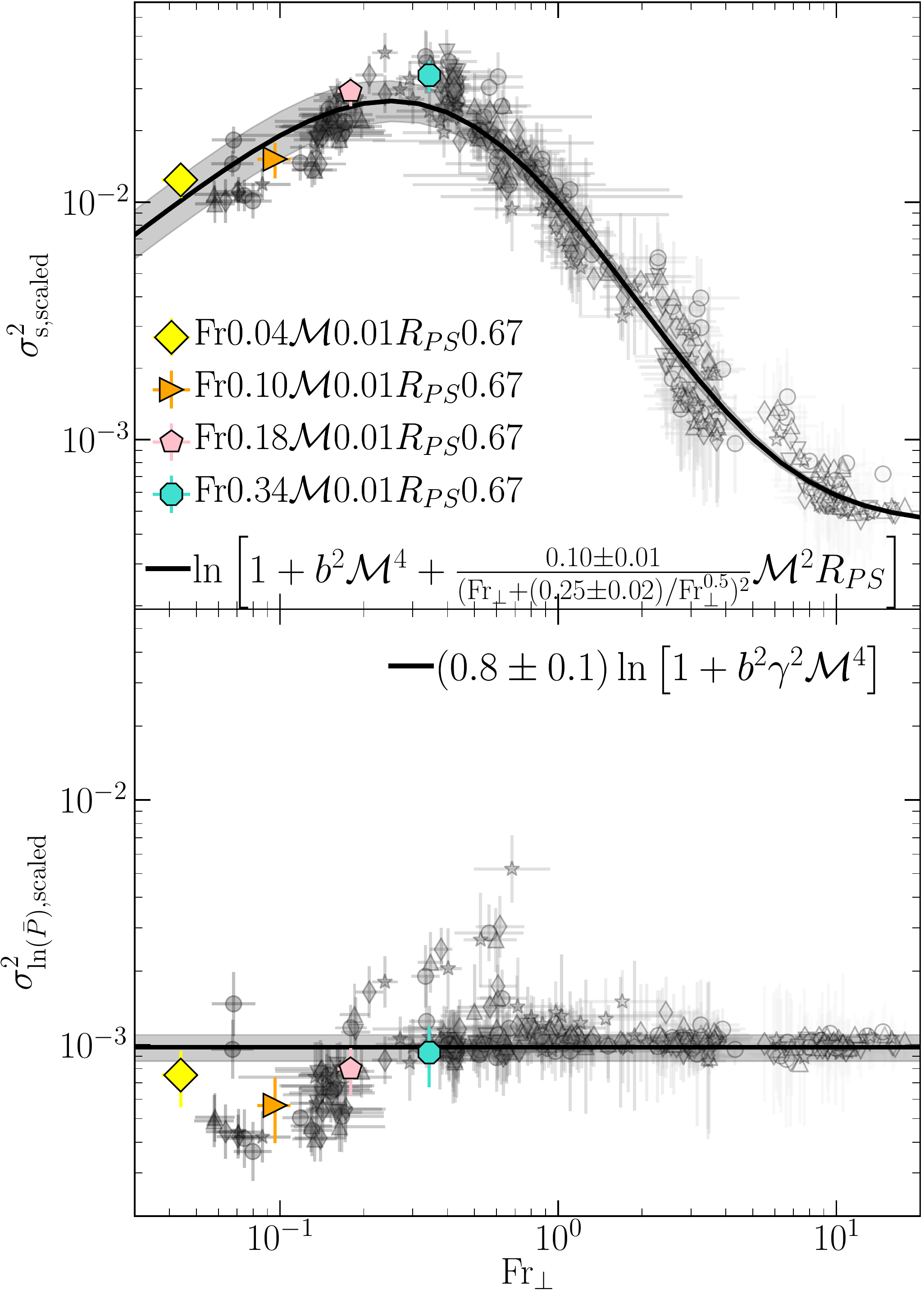}	
	\caption[sigma-Fr-relation at low mach]{The scaled density ($\sigma_{s\mathrm{,scaled}}^2$; upper panel) and pressure fluctuations ($\sigma_{\log{\bar{P}\mathrm{,scaled}}}^2$; lower panel) for our simulations with $\mathcal{M}\approx0.01$, $R_{PS}=0.67$, and $\mathrm{Fr}_\perp\lesssim0.4$. We also show our data points from \cref{fig:sig-oneMach-Fr} in grey. The low-$\mathcal{M}$ data follows our proposed scaling relation for both density and pressure fluctuations.}
	\label{fig:sig-scaled-Fr-mach001}
\end{figure}

In this subsection we test the scaling of the velocity ratio, density and pressure fluctuations with $\mathrm{Fr}_\perp$ through 4~simulations at $\mathcal{M}\approx0.01$, $R_{PS}=0.67$ and $\mathrm{Fr}_\perp\lesssim0.4$. See \cref{tab:lowMach_sim_params} for the simulation parameters. Since these simulations have $H_P=H_\rho$, the sound speed is uniform throughout the domain and the turbulent pressure is negligible compared to the thermal pressure (turbulent pressure $\sim10^{-4}$ thermal pressure). Since $H_P,H_\rho>1.0$, the density and pressure gradients are not very steep. Thus, these simulations do not suffer from %any of the instabilities 
fluctuations near the boundaries mentioned in \cref{subsubsec:BC,subsubsec:slab_division}.

In \cref{fig:vz-vperp-Fr-mach001} we show the ratio between $v_z$ and $v_\perp$ for these simulations. The ratio follows the $\mathrm{Fr}^{0.7}$ scaling. In \cref{fig:sig-scaled-Fr-mach001} we show $\sigma^2_{s,\mathrm{scaled}}$ and $\sigma^2_{\ln(\bar{P}),\mathrm{scaled}}$, and we find that they also follow the scaling relation in the low-$\mathrm{Fr}_\perp$ regime---$\sigma^2_{s,\mathrm{scaled}}$ decreases with decreasing $\mathrm{Fr}_\perp$ and $\sigma^2_{\ln(\bar{P}),\mathrm{scaled}}$ is independent of $\mathrm{Fr}_\perp$.

\subsubsection{Comparing the density scaling relation with other studies}\label{subsubsec:fitting_from other studies}

\begin{figure}
		\centering
	\includegraphics[width=0.99\columnwidth]{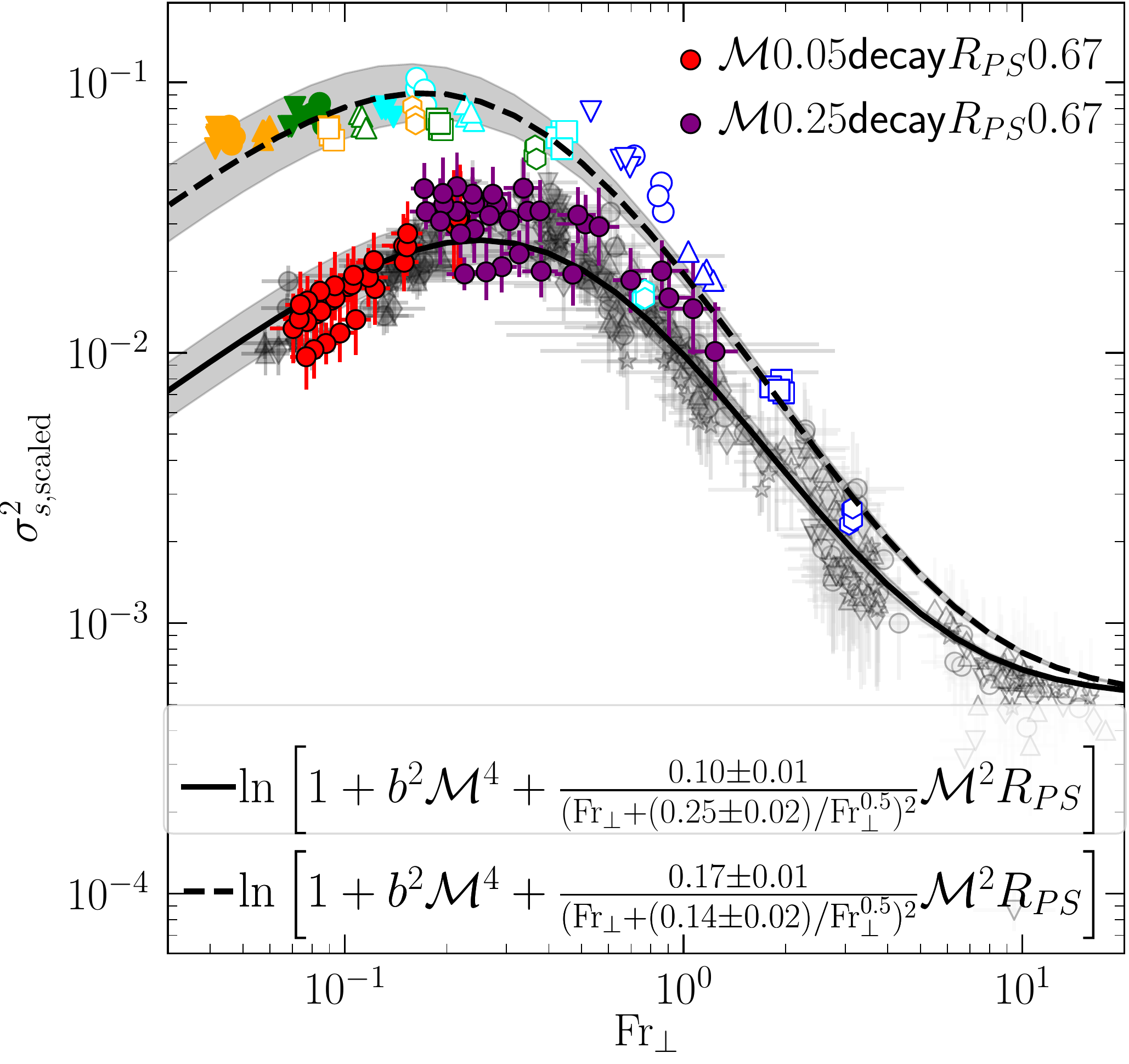}	
	\caption[Shi-sigs-Fr-plots]{The scaled density fluctuations ($\sigma_{s,\mathrm{scaled}}^2$) vs $\mathrm{Fr}_\perp$ with the coloured (cyan, green and orange) datapoints taken from \citetalias{Shi2019} and scaled (see the main text). We follow the same colour and marker scheme as in figure~8 of \citetalias{Shi2019}, where different colours represent different values of $N$ (which correspond to simulations with different levels of stratification). The open symbols are for $t\leq2/N$ and the filled symbols are for $t>2/N$. We show our data points from the top panel of \cref{fig:sig-oneMach-Fr} in grey. We have also plotted data points from our turbulence decay simulations in red and purple. The dashed line shows the best fit to \citetalias{Shi2019}'s data and the solid line shows the best fit to our data. There seems to be a small scaling discrepancy which becomes larger at smaller Fr$_\perp$. We discuss more about this discrepancy and its possible sources in the main text. 
	}
	\label{fig:Shi-sigs-scaled-Fr}
\end{figure}

In this subsection, we test our scaling relation against data from \citetalias{Shi2019}, who study  decaying turbulence in a stratified medium, using a setup similar to ours. We have received the simulation snapshots from \citeauthor{Shi2019} and applied our analysis methods to calculate $\sigma_s^2$ and $\mathrm{Fr}_\perp$.\footnote{Our values of $\mathrm{Fr}_\perp$ for \citetalias{Shi2019} differ by a factor of $2\pi$. This results from the slightly different definitions of the integral scale $l_\perp$ (see equation~3 in \citetalias{Shi2019}), where $2\pi$ is the conversion factor from wavenumber space to real space.} We scale these quantities to $\sigma_{s,\text{scaled}}^2$ with $\mathcal{M}=0.25$ and $R_{PS}=2.33$, using \cref{eq:sigs_scaled}. 

In \cref{fig:Shi-sigs-scaled-Fr}, we show $\sigma_{s,\text{scaled}}^2$ as a function of $\mathrm{Fr}_\perp$, using the same colour scheme (cyan, green and orange) and markers as figure~8 in \citetalias{Shi2019}. We have greyed out our data points in the background. \citetalias{Shi2019} study decaying turbulence, and therefore all variables are time ($t$) dependent in their study. The open symbols in \cref{fig:Shi-sigs-scaled-Fr} are for $t\leq2/N$ and the filled symbols are for $t>2/N$. The dashed line is the best fit to the scaled \citetalias{Shi2019} data and the solid line is the best fit to our data. 

We find that at high $\mathrm{Fr}_\perp$, $\sigma_{s,\mathrm{scaled}}^2$ for \citetalias{Shi2019}'s data shows a similar dependence on $\mathrm{Fr}_\perp$ as our scaling relation predicts. However, the exact dependence of $\sigma_{s,\mathrm{scaled}}^2$ on $\mathrm{Fr}_\perp$ is somewhat different, i.e., we see that $\sigma_{s,\mathrm{scaled}}^2$ is higher by almost a factor of $2$--$4$ for $\mathrm{Fr}_\perp\lesssim0.5$ becomes independent of $\mathrm{Fr}_\perp$ for $\mathrm{Fr}_\perp\lesssim0.3$. We find that the fitting parameters $\zeta_1^2$ and $\zeta_2$ (see \cref{eq:sigs_combined}) take the values $0.17\pm0.01$ and $0.14\pm0.02$ respectively. 

The open data points for $t\leq2/N$ in \citetalias{Shi2019} correspond to the moderate-weakly stratified turbulence regime ($\mathrm{Fr}_\perp\gtrsim0.5$). In this limit,
\begin{equation}\label{eq:moderate_strat_delrho}
    \delta\bar{\rho}^2\approx \delta\bar{\rho}_{\text{buoy}}^2\propto\mathcal{M}^2/\mathrm{Fr}_\perp^2=N^2\ell_\perp^2/c_s^2.
\end{equation}
For a given stratification profile (fixed $H_P$, $H_\rho$ and corresponding $N$) and $t\leq2/N$, $\delta\bar{\rho}^2$ is independent of $\mathcal{M}$, as seen in figure~8 of \citetalias{Shi2019}. This has significant implications for observational studies, which infer turbulent velocities from surface brightness fluctuations, which are in turn caused by density fluctuations.
For $t>2/N$, $\mathrm{Fr}_\perp\lesssim0.5$, where the $\sigma_s^2$ dependence on $\mathrm{Fr}_\perp$ is weak near the peak of the proposed scaling relation and hence $\delta\bar{\rho}^2\propto\mathcal{M}^2$.

In order to directly compare against \citetalias{Shi2019}, we conduct two sets of turbulence decay simulations with $\mathcal{M}_{\text{bin}}=0.05$ and $0.25$, $H_P=H_\rho=1.0$ with grid resolution $512^2\times768$. Once turbulence reaches a roughly steady state (after $\approx3 t_{\mathrm{eddy}}$), we switch off external driving. We simulate five statistically similar instances of decaying turbulence by changing the random number seed for the turbulent driving.

We calculate the average value of $\sigma_s^2$ over time intervals of $2 t_{\text{eddy}}$ in the decay phase. We further average these values across the five runs. The red and purple circles in \cref{fig:Shi-sigs-scaled-Fr} represent $\sigma_{s,\text{scaled}}^2$ for these decaying turbulence simulations. Both the purple and red data points agree with our proposed scaling relation. Unlike \citetalias{Shi2019}'s data, for $\mathrm{Fr}_\perp\lesssim0.3$, $\sigma^2_{s,\text{scaled}}$ in decaying turbulence simulations decreases with decreasing $\mathrm{Fr}_\perp$, similar to our steady-state turbulence runs.

The differences in the values of $\sigma_{s,\text{scaled}}^2$ at low $\mathrm{Fr}_\perp$ may arise due to the spurious sound waves as discussed in \cref{subsubsec:BC}. Due to the reflective boundary conditions along the $z$ direction, they can form standing waves in pressure (X.~Shi, private communication). If the amplitude of density fluctuations due to the standing waves is comparable to $\delta\bar{\rho}_{\mathrm{buoy}}$, then the net density fluctuations would not show a decrease at the low $\mathrm{Fr}_\perp$ limit. Our turbulence decay simulations do not suffer from this because of the new Dirichlet boundary conditions that we apply to the density and pressure. These simulations are at low rms Mach number and use an initially isothermal stratification profile ($H_P=H_\rho=1.0$). Hence they also do not suffer from the numerical fluctuations at the upper boundary discussed in \cref{subsubsec:slab_division}.

\subsection{Density fluctuations versus velocity power spectra in strongly stratified turbulence}\label{subsec:etak}

\begin{figure}
		\centering
	\includegraphics[width=0.99\columnwidth]{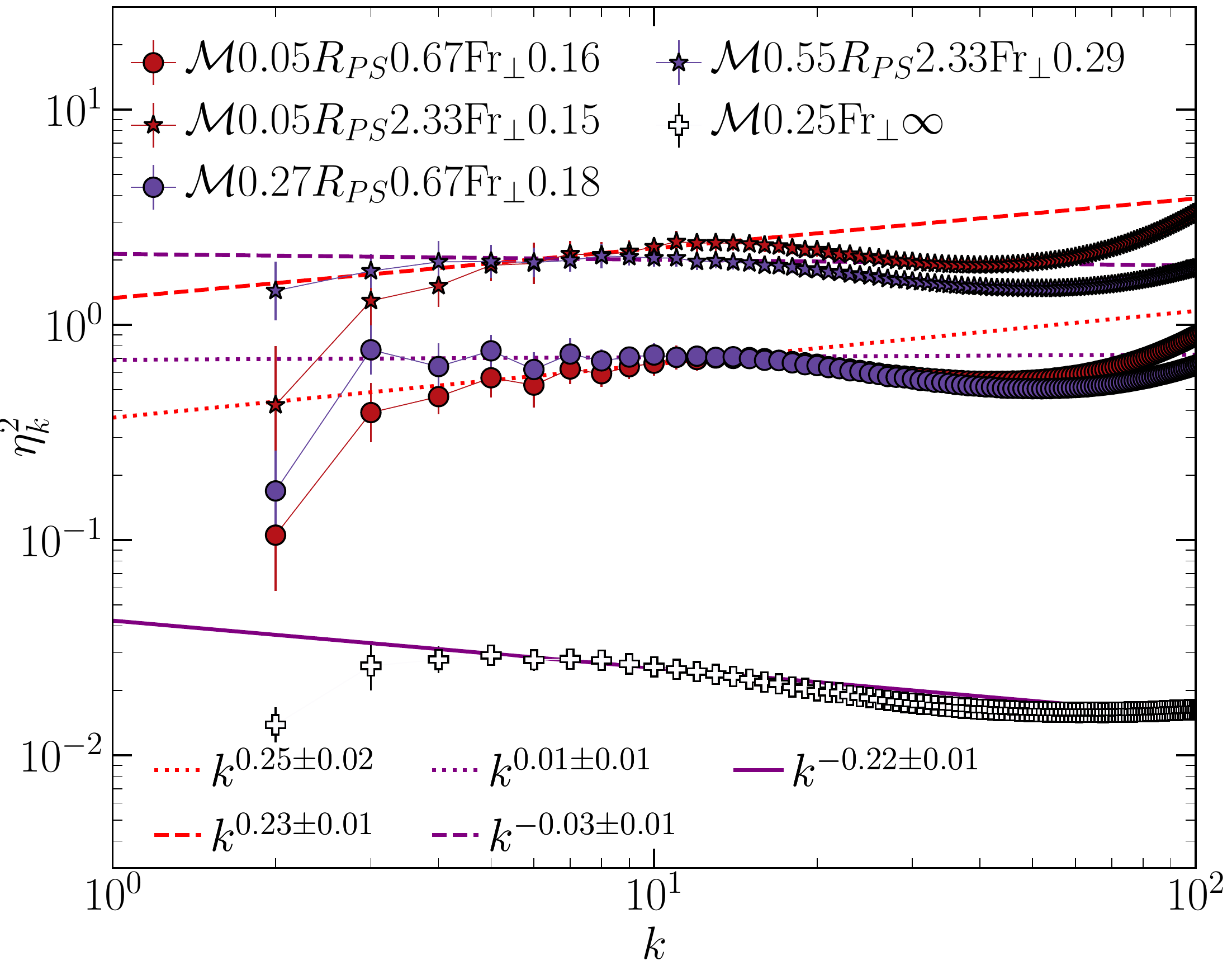}	
	\caption[eta-lowFr]{The ratio of density fluctuations to velocity power spectra, $\eta_k^2$, defined in \cref{eq:etak}, as a function of the wavenumber $k$, for our high-resolution, strongly stratified simulations. In this strongly stratified limit (which does not necessarily apply to the ICM in general; see Fig. 1 in \citetalias{Mohapatra2020}), $\eta_k^2$ depends strongly on $R_{PS}$ and weakly on $k$, $\mathcal{M}$ and $\mathrm{Fr}_\perp$.}
	\label{fig:etak-lowFr}
\end{figure}

\begin{figure}
		\centering
	\includegraphics[width=0.99\columnwidth]{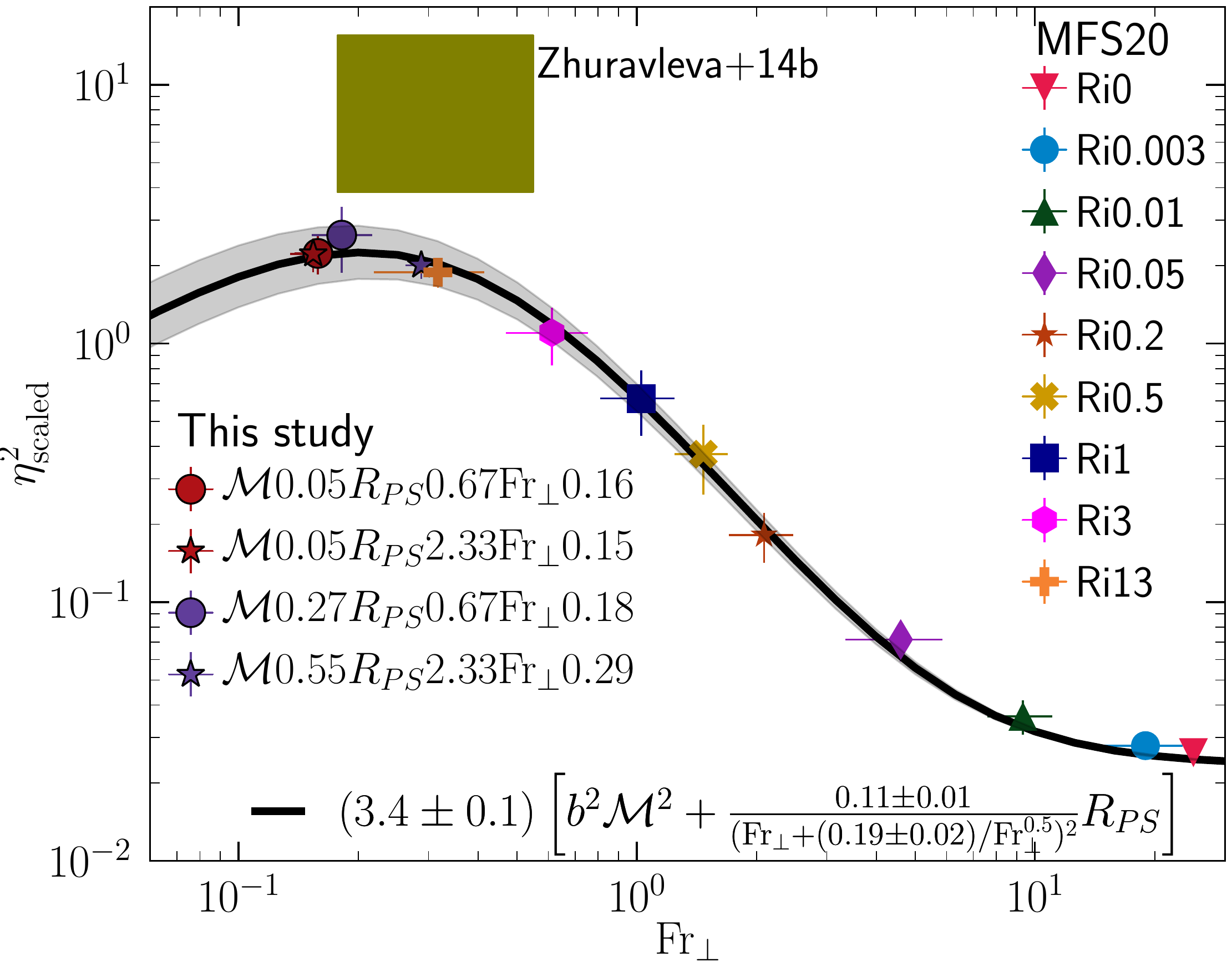}	
	\caption[etasq-Fr]{The scaled density to velocity fluctuation ratio $\left(\eta^2_{\text{scaled}}\right)$ as a function of $\mathrm{Fr}_\perp$ for the high-resolution simulations from this study and from \citetalias{Mohapatra2020}. The ratio  ($\eta_k^2$) is averaged over the inertial range of turbulence ($6\leq k \leq 15$). The fit (solid line) shown is plotted using $\mathcal{M}=0.25$ and $R_{PS}=2.33$, corresponding to the parameters used in \citetalias{Mohapatra2020}. The ratio ($\eta_k^2$) roughly scales as $\sigma_{\bar{\rho}}^2/\mathcal{M}^2$ and is independent of $\mathcal{M}$ in the strongly stratified turbulence limit. We also show the values of $\eta^2_{\text{scaled}}$ and $\mathrm{Fr}$  clusters from \cite{zhuravleva2014relation} in the olive coloured shaded region. The values seem to be slightly larger than predicted by our scaling relation at low $\mathrm{Fr}_\perp$. We discuss this discrepancy in the main text.}
	\label{fig:etasq-Fr}
\end{figure}

In the previous subsections we have investigated the scaling of rms density fluctuations with the rms Mach number, $\mathrm{Fr}_\perp$, and $R_{PS}$. Another quantity relevant for observations is the amplitude and spectral scaling of density fluctuations and velocity power spectra. This is because turbulent velocities are more difficult to measure directly in observations, and often the surface brightness fluctuations are used to infer properties of the turbulence (\citealt{zhuravleva2014turbulent,Zhuravleva2015MNRAS,zhuravleva2018} and \citealt{Simionescu2019SSRv} for a review).

The ratio between Fourier one-component density fluctuation amplitudes $\delta\bar{\rho}_k$ and Fourier one-component velocity amplitudes $v_{1,k}$ is denoted by $\eta_k$ and it is related to the density fluctuations and velocity power spectra by
\begin{equation}
    \eta_k^2=(\delta\bar{\rho}_k)^2/(v_{1,k}/c_s)^2\approx3P(\delta\bar{\rho}_k)/P(\mathcal{M}_k),\label{eq:etak}
\end{equation}
where $P(\delta\bar{\rho}_k)$ and $P(\mathcal{M}_k)$ are the density fluctuations and velocity power spectra. Density and velocity are normalised with respect to the stratification profile and the speed of sound, respectively.
The ratio given by $\eta_k$ is an important parameter for the observational studies listed above. The Fourier transform of the normalised X-ray SB fluctuations $\mathrm{\bar{SB}}_k$ are used to calculate $\bar{\rho}_k$ and then $v_{1,k}$ is calculated using \cref{eq:etak}.

All the observational studies we mentioned above assume $\eta_k^2\approx1$, independent of $k$, $\mathcal{M}$ and $\mathrm{Fr}_\perp$. A value of $\eta_k^2\approx1$ was calibrated using cosmological simulations in \cite{zhuravleva2014relation} and large cluster scale (1000~kpc) simulations in \cite{gaspari2014}. In \cite{Mohapatra2019}, we found $\eta_k^2\propto\mathcal{M}^2$ through idealised box simulations of unstratified turbulence. We also noted that the amplitude of $\eta_k^2$ increases by two orders of magnitude on including cooling. 
In subsection~3.9.3 of \citetalias{Mohapatra2020}, we studied the variation of $\eta_k^2$ for weakly and moderately stratified turbulence ($\mathrm{Fr}_\perp\gtrsim 0.5$, $\mathcal{M}\approx0.25$ and $R_{PS}=2.33$). We showed that $\eta_k^2$ increases with increasing stratification and reaches $\eta_k^2\approx1$ for $\mathrm{Fr}_\perp\approx0.7$. Here we have extended this study to the strongly stratified turbulence regime (up to $\mathrm{Fr}_\perp\approx0.1$).

In \cref{fig:etak-lowFr}, we show $\eta_k^2$ as a function of wavenumber $k$ for our high-resolution, strongly stratified simulations. For reference, we show $\eta_k^2$ for the unstratified turbulence run (Ri0, $\mathrm{Fr}_\perp\xrightarrow{}\infty$) from \citetalias{Mohapatra2020}. The quantity $\eta_k^2$ is relatively flat with respect to $k$, for all of these simulations, which implies that $P(\delta\bar{\rho}_k)$ and $P(\mathcal{M}_k)$ follow similar spectral scaling.

We also show $\eta^2$, the amplitude of $\eta_k^2$ averaged over $6\leq k\leq15$ as a function of $\mathrm{Fr}_\perp$ in \cref{fig:etasq-Fr} for the high-resolution simulations ($1024^2\times1536$) of this study and \citetalias{Mohapatra2020}\footnote{All simulations in \citetalias{Mohapatra2020} had $R_{PS}=2.33$ and $\mathcal{M}\sim0.25$.}. For the two simulations with $R_{PS}\neq2.33$, we show $\eta^2_{\text{scaled}}=\eta^2\times(2.33/R_{PS})$. We fit the data using the expression for $\sigma_{\bar{\rho}}^2/\mathcal{M}^2$ from \cref{eq:delrho_combined} and $\mathcal{M}=0.25$ and $R_{PS}=2.33$, with $\zeta_1$ and $\zeta_2$ as fit parameters. The fitted factor of $(3.4\pm0.1)$ is close to the expected factor of $3$ from the 3D to 1D conversion; see \cref{eq:etak}. We have also plotted $\eta^2_\text{scaled}$ for cluster simulations from \cite{zhuravleva2014relation} in the olive coloured shaded region in \cref{fig:etasq-Fr}, which are slightly higher than our values and scaling relation. The differences in values of $\eta^2_{\text{scaled}}$ may arise from the choice of averaging regions, in this case $500~\mathrm{kpc}$ regions in the clusters of \citealt{zhuravleva2014relation}. The density fluctuations in their study are also calculated as deviations from a beta-profile fit to these $500~\mathrm{kpc}$ regions and thus are expected to be higher than our instantaneous $z$-slice averaging (described in \cref{subsec:turb-params}). In addition to this, since turbulence in these cosmological simulations is driven more naturally by galaxy mergers and in-falls, they may include compressive components (which are excluded in our simulations), increasing the value of the turbulence driving parameter $b$ and generating larger density fluctuations. The cosmological simulations may also suffer from a lack of resolution on small scales and have a higher numerical viscosity compared to our idealised box simulations.

The amplitude of $\eta_k^2$ has the same dependence on $\mathcal{M}$, $\mathrm{Fr}_\perp$, and $R_{PS}$, as the ratio of the rms fluctuations
$\delta\bar{\rho}^2/\mathcal{M}^2$. Since $\delta\bar{\rho}_{\text{buoy}}^2$ is the main component of  $\delta\bar{\rho}^2$ for $\mathrm{Fr}_\perp\approx0.1$, $\eta^2$ scales similar to $\delta\bar{\rho}_{\text{buoy}}^2/\mathcal{M}^2$. Hence $\eta^2$ is roughly independent of $\mathcal{M}$, weakly dependent on $\mathrm{Fr}_\perp$ (since $\sigma_s$ versus $\mathrm{Fr}_\perp$ peaks at around $\mathrm{Fr}_\perp\approx0.1$), and linearly dependent on $R_{PS}$, in agreement with \cref{eq:delrho_buoy_combined}.

Combining the results of this work with those of \citet{Mohapatra2019} and \citetalias{Mohapatra2020}, we have studied the variation of $\eta_k^2$ with $\mathcal{M}$, $\mathrm{Fr}_\perp$, and $R_{PS}$, scanning the parameter space relevant for the ICM. We find that the ratio $\eta_k^2$ is almost invariant with the wavenumber $k$\footnote{Except for the heating and cooling simulations in \cite{Mohapatra2019}, where switching on radiative cooling led to a steepening.}. The amplitude of $\eta_k^2$ varies as $\delta\bar{\rho}^2/\mathcal{M}^2$. 
Based on \cref{eq:delrho_combined} and the fit in \cref{fig:etasq-Fr}, we propose a new scaling relation for $\eta^2$:
\begin{equation}
    \eta^2=\left(3.4\pm0.1\right) \left[b^2\mathcal{M}^2+\frac{(0.11\pm0.01)R_{PS}}{\left(\mathrm{Fr}_\perp+(0.19\pm0.02)/\sqrt{\mathrm{Fr}_\perp}\right)^2}\right]. \label{eq:etak_overall_scaling}
\end{equation}
For $\mathrm{Fr}_\perp\lesssim1$, $\eta^2$ approaches 1 and is independent of $\mathcal{M}$, which was the limit studied in \cite{zhuravleva2014relation} and \cite{gaspari2014}. However, as we showed in figure~1 of \citetalias{Mohapatra2020} (also in figure~2 of \citetalias{Shi2019}), the ICM can have $\mathrm{Fr}_\perp$ between $0.1$--$100$. Thus, we suggest that observational studies use \cref{eq:etak_overall_scaling} to obtain a more accurate estimate of turbulent velocities from density (surface brightness) fluctuations. Since $\mathrm{Fr}_\perp\lesssim1$ also marks the onset of large-scale density anisotropy (see lower panels of fig.~\ref{fig:dens-proj-2d} and also fig.~10 in \citetalias{Mohapatra2020}), one could use the peak value of $\eta^2$, where it varies slowly with $\mathrm{Fr}_\perp$ for relating the ICM density and velocity fluctuations.
\FloatBarrier

\section{Caveats and future work}\label{sec:caveats-future}
In this section we discuss possible shortcomings of our work and possible methods to address some of them. 

The buoyancy Reynolds number $\Re_{\text{buoy}}$ for stratified turbulence is given by $\mathrm{Re}_\perp\mathrm{Fr}_\perp^2$, where $\mathrm{Re}_\perp$ is the turbulent Reynolds number in the transverse direction. When $\Re_{\text{buoy}}\gg1$, viscous forces can be ignored on the integral scale $\ell_\perp$ \citep{davidson_2013}.
The ratio $\mathrm{Re}_\perp$ scales as $n^{4/3}$ in numerical simulations where $n$ is the number of resolution elements \citep{Frisch1995,Haugen2004PhRvE,Benzi2008PhRvL,Federrath2011ApJ}. Hence for low $\mathrm{Fr}_\perp$ simulations, $\Re_{\text{buoy}}\approx4096\times\mathrm{Fr}_\perp^2\approx10$ for $\mathrm{Fr}_\perp0.05$, using $n=512$. Thus, we approach the limit $\Re_{\text{buoy}}\approx1$ for our strongly stratified simulations, implying that viscous forces are almost of the same order as the inertial forces on the integral scale for $\mathrm{Fr}_\perp\approx0.05$. Using higher grid resolution is one of the ways to avoid this issue.

In \cref{subsubsec:slab_division}, we mentioned that the positive $z$ boundaries (at $z=0.75$) are unstable for steep pressure profiles ($H_P<0.25$). This happens because for the steepest pressure profiles with $H_P\approx0.1$, the initial pressure varies by more than a factor of $10^6$ across the entire domain. These steep gradients could possibly make the code numerically unstable near the low-pressure positive $z$ boundary, especially when turbulent pressure becomes comparable to the thermal pressure. Using smaller box sizes (so that pressure values at the positive box boundary are higher) and/or higher resolution along the $z$ direction is a possible solution to this problem. This method has been used by many fluid mechanics studies to study strongly stratified turbulence \citep{Lindborg2006,Brethouwer2008}. 

In this paper, we have only studied solenoidally-driven stratified turbulence. The driving parameter $b$ is also part of the scaling relation \citep{Federrath2008,federrath2010}. Compressively-driven turbulence is expected to generate larger density fluctuations and its effect is supposed to be captured by the driving parameter $b$, which we have not varied in any of our simulations. In nature, turbulence is supposed to be a mixture of compressive and solenoidal mode. Galaxy infall and mergers could possibly drive some compressive modes \citep{Churazov2003,Federrath2017IAUS} in the ICM. Effects of different levels of compressively-driven turbulence would be an interesting follow-up study.

We have ignored the effects of radiative cooling, thermal conduction, and magnetic fields in this work. Radiative cooling \citep{Mohapatra2019,Grete2020ApJ} and thermal conduction \citep{gaspari2013constraining,gaspari2014} have been shown to affect the $\delta\bar{\rho}$--$\mathcal{M}$ scaling relation. However, this additional physics is beyond the scope of the current paper and will be addressed in a follow-up study.

\section{Conclusions}\label{sec:Conclusion}
We have studied the parameter space relevant to subsonic, stratified turbulence in the ICM through idealised high-resolution hydrodynamic simulations. We have covered the parameter regime $0.05<\mathrm{Fr}_\perp<12.0$, $0.05\leq\mathcal{M}\leq0.4$ and $0.33\leq R_{PS}\leq 2.33$ through 96 simulations. The main results of this study are as follows:
\begin{enumerate}

    \item We have extended the scaling relation between the rms density fluctuations (denoted in log-scale by $\sigma_s$), the rms Mach number ($\mathcal{M}$), the perpendicular Froude number ($\mathrm{Fr}_\perp$), and the ratio between pressure and entropy scale heights ($R_{PS}$) to the strong stratification limit ($\mathrm{Fr}_\perp\ll1$). The new scaling relation is $\sigma_s^2=\ln\left[1+b^2\mathcal{M}^4+\frac{0.10\mathcal{M}^2R_{PS}}{\left(\mathrm{Fr}_\perp+0.25/\sqrt{\mathrm{Fr}_\perp}\right)^2}\right].$ The density fluctuations increase with decreasing $\mathrm{Fr}_\perp$ for $\mathrm{Fr}_\perp\gtrsim0.2$, saturate and then decrease slowly for $\mathrm{Fr}_\perp$. We have shown that the density fluctuations in all of our 100 simulations follow this scaling relation. Our results also qualitatively agree with the turbulence decay study in \citetalias{Shi2019}.

    \item We have also extended the scaling of pressure fluctuations to the limit $\mathrm{Fr}_\perp\ll1$. We find that $\sigma_{\ln\bar{P}}$ is independent of the stratification parameters and depends only on $\mathcal{M}$. The scaling relation remains  $\sigma_{\ln\bar{P}}=\ln\left[1+b^2\gamma^2\mathcal{M}^4\right]$. Since pressure fluctuations are unaffected by stratification, they can be used to obtain fairly accurate estimates of turbulent velocities through tSZ observations.
    
    \item The ratio $\eta_k$ between normalised one-dimensional Fourier density ($\bar{\rho}_k$) and velocity amplitudes ($v_{1,k}/c_s$) approximately scales as $\sigma_{\bar{\rho}}/\mathcal{M}$ and saturates to $\approx1$ at $\mathrm{Fr}_\perp\approx0.2$. The square of its amplitude is given by
    $ \eta^2=3.4 \left[b^2\mathcal{M}^2+\frac{0.11R_{PS}}{\left(\mathrm{Fr}_\perp+0.19/\sqrt{\mathrm{Fr}_\perp}\right)^2}\right]$.

    \item The ratio between velocity components parallel and perpendicular to gravity ($v_z/v_\perp$) is roughly constant with respect to $\mathrm{Fr}_\perp$ for $\mathrm{Fr}_\perp\gtrsim1$, as expected for weak stratification. For $0.5\lesssim\mathrm{Fr}_\perp\lesssim2$, the ratio weakly decreases with decreasing $\mathrm{Fr}_\perp$, and for $\mathrm{Fr}_\perp\lesssim0.5$, it varies as $\mathrm{Fr}_\perp^{0.7}$.

\end{enumerate}

\section*{Acknowledgements}
This work was carried out during the ongoing COVID-19 pandemic. The authors would like to acknowledge the health workers all over the world for their role in fighting in the frontline of this crisis. We thank the anonymous referee for their useful comments, which helped improve this work. RM acknowledges helpful discussions with Piyush Sharda. RM thanks Xun Shi and Irina Zhuravleva for providing data for subsection~\ref{subsubsec:fitting_from other studies} and \cref{subsec:etak}, respectively and useful discussions. CF acknowledges funding provided by the Australian Research Council (Discovery Project DP170100603 and Future Fellowship FT180100495), and the Australia-Germany Joint Research Cooperation Scheme (UA-DAAD). PS acknowledges a Swarnajayanti Fellowship from the Department of Science and Technology, India (DST/SJF/PSA-03/2016-17). We further acknowledge high-performance computing resources provided by the Leibniz Rechenzentrum and the Gauss Centre for Supercomputing (grants~pr32lo, pr48pi and GCS Large-scale project~10391), the Australian National Computational Infrastructure (grant~ek9) in the framework of the National Computational Merit Allocation Scheme and the ANU Merit Allocation Scheme. The simulation software FLASH was in part developed by the DOE-supported Flash Center for Computational Science at the University of Chicago.
%%%%%%%%%%%%%%%%%%%%%%%%%%%%%%%%%%%%%%%%%%%%%%%%%%

\section{Data Availability}
All the relevant data associated with this article is available upon request to the corresponding author.

\section{Additional Links}
Movies of projected density and density fluctuations of different simulations are available at the following links on youtube:
\begin{enumerate}
    \item \href{https://youtu.be/arvouYKQ5Ic}{Movie} of the representative simulations used in  \cref{fig:dens-proj-2d}.
    \item \href{https://youtu.be/6bHH9FHPUq0}{Movie} of low $\mathcal{M}$, low $\mathrm{Fr}_\perp$ simulations listed in \cref{tab:lowMach_sim_params}.
    \item \href{https://youtu.be/W3q3izzD3Zo}{Movie} of two sample stratified turbulence decay simulations used in \cref{subsubsec:fitting_from other studies}.
\end{enumerate}



\bibliographystyle{mnras}
\bibliography{refs.bib} % if your bibtex file is called example.bib

%%%%%%%%%%%%%%%%%%%%%%%%%%%%%%%%%%%%%%%%%%%%%%%%%%

%%%%%%%%%%%%%%%%% APPENDICES %%%%%%%%%%%%%%%%%%%%%

\appendix
\renewcommand\thefigure{\thesection A\arabic{figure}} 
\setcounter{figure}{0}   
\setcounter{table}{0}   
\section*{Appendix A}\label{app:sigma_rho}

\begin{figure}
		\centering
	\includegraphics[width=0.99\columnwidth]{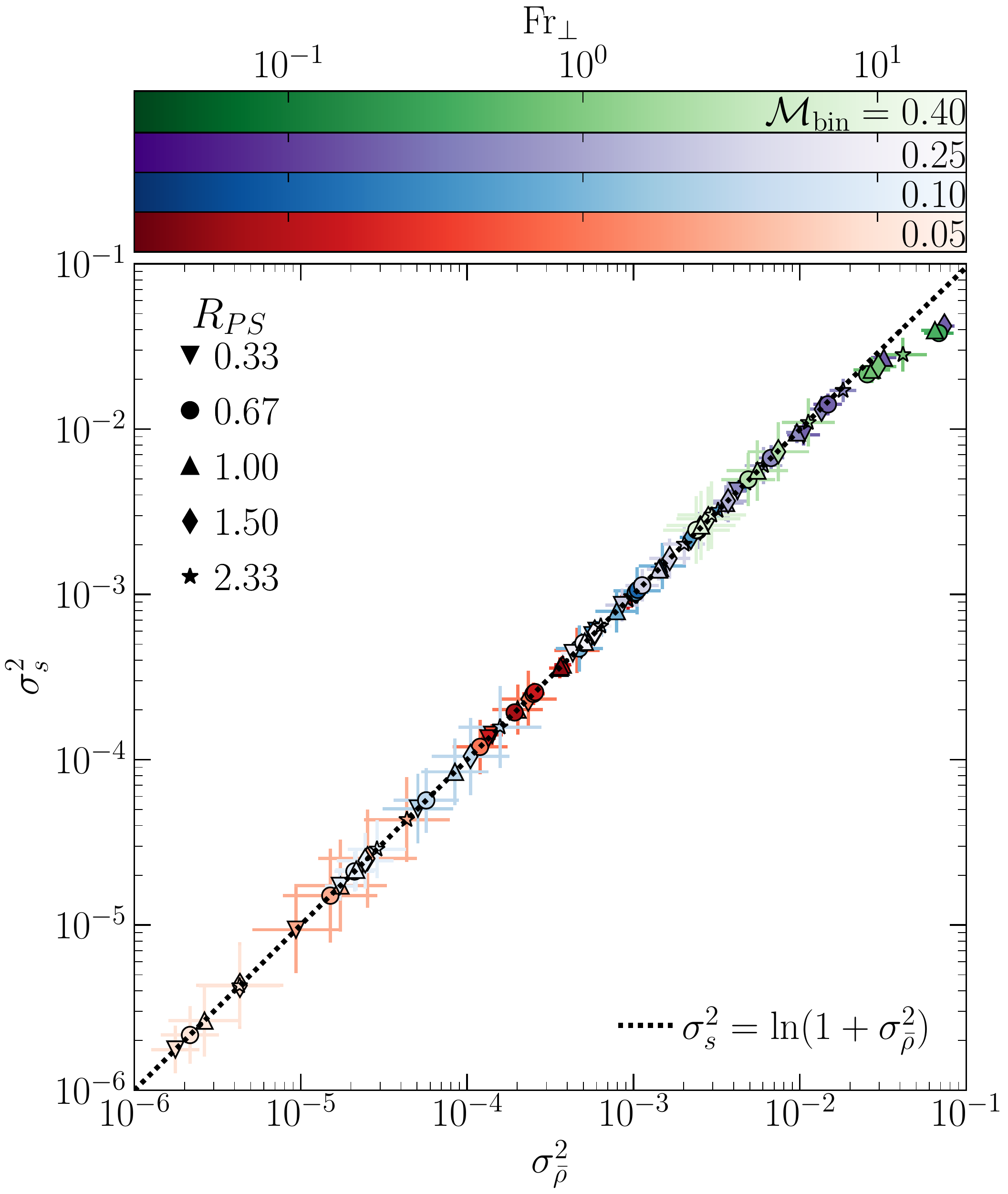}	
	\caption[sigs-nrho]{The spread in $\ln \rho$ ($\sigma_s^2$) as a function of the width in $\rho$ ($\sigma_{\bar{\rho}}^2$) for all our runs.  The symbol shape corresponds to $R_{PS}$, the colour to the Mach number, and the shading to Fr$_\perp$. The data is fit well by $\sigma_s^2=\ln(1+\sigma_{\bar{\rho}}^2)$, as discussed in the last paragraph of \cref{subsubsec:dens_pres_fluc_Froude}.}
	\label{fig:sigs-signrho}
\end{figure}

\Cref{fig:sigs-signrho} shows the relation between $\sigma_s^2$ and $\sigma_{\bar{\rho}}^2$. The relation $\sigma_s^2=\ln(1+\sigma_{\bar{\rho}}^2)$ holds for log-normal distributions of density and/or small density fluctuations ($\sigma_{\bar{\rho}}\ll1$) \citep{Price2011ApJ}. We use this relation to explain the scaling of $\sigma_s^2$ based on the turbulent and buoyant components of $\sigma_{\bar{\rho}}^2$ in our study.

\FloatBarrier
\label{sec:Appendix}

\onecolumn

\section*{Appendix B}\label{app:simtab}\label{sec:appendix_table}
\renewcommand\thetable{\thesection B\arabic{table}} 

\Cref{tab:sim_list} provides a long version of \cref{tab:sim_params}, listing all the simulations, including all input parameters and relevant calculated/output variables.
\begin{center}

\def\arraystretch{1.35}
\setlength{\tabcolsep}{11.2pt}
\setcounter{table}{0}   
\begin{longtable}{lccccccc}\caption{Full list of simulations}\label{tab:sim_list}\\
    \hline
    $\mathcal{M}_{\text{bin}}$ & $H_\rho$ & $H_P$ & Resolution & $\mathrm{Fr}_\perp$ & $\mathcal{M}$ & $\sigma_s^2$ & $\sigma_{\ln\bar{P}}^2$\\
    \hline\\
    \endfirsthead
    \multicolumn{8}{@{}l}{\ldots continued}\\\hline
    $\mathcal{M}_{\text{bin}}$ & $H_\rho$ & $H_P$ & Resolution & $\mathrm{Fr}_\perp$ & $\mathcal{M}$ & $\sigma_s^2$ & $\sigma_{\ln\bar{P}}^2$\\
    \hline\\
    \endhead
    \multicolumn{8}{r@{}}{continued \ldots}\\
    \endfoot
    \hline
    \endlastfoot
    $0.01$ & $8.00$ & $8.00$ & $512^2\times768$ & $3.4^{-0.3}_{+0.3}\times10^{-1}$ & $0.0087\pm0.0003$ & $1.1^{-0.1}_{+0.2}\times10^{-5}$ & $1.4^{-0.3}_{+0.4}\times10^{-9}$ \\
    $0.01$ & $4.00$ & $4.00$ & $512^2\times768$ & $1.8^{-0.2}_{+0.2}\times10^{-1}$ & $0.0089\pm0.0005$ & $1.1^{-0.1}_{+0.2}\times10^{-5}$ & $1.4^{-0.3}_{+0.3}\times10^{-9}$ \\
    $0.01$ & $2.00$ & $2.00$ & $512^2\times768$ & $9.4^{-1.3}_{+1.5}\times10^{-2}$ & $0.0098\pm0.0008$ & $7.4^{-1.2}_{+1.5}\times10^{-6}$ & $1.6^{-0.4}_{+0.6}\times10^{-9}$ \\
    $0.01$ & $1.00$ & $1.00$ & $512^2\times768$ & $4.4^{-0.6}_{+0.7}\times10^{-2}$ & $0.010\pm0.001$ & $6.4^{-1.1}_{+1.4}\times10^{-6}$ & $2.5^{-0.6}_{+0.8}\times10^{-9}$ \\
    \hline
    $0.05$ & $17.32$ & $13.86$ & $256^2\times384$ & $9.5^{-2.2}_{+2.9}\times10^{0}$ & $0.053\pm0.005$ & $1.7^{-0.5}_{+0.8}\times10^{-6}$ & $2.0^{-0.4}_{+0.5}\times10^{-6}$ \\
    $0.05$ & $5.48$ & $4.38$ & $256^2\times384$ & $2.7^{-0.6}_{+0.9}\times10^{0}$ & $0.053\pm0.005$ & $9.1^{-4.3}_{+8.1}\times10^{-6}$ & $2.0^{-0.5}_{+0.6}\times10^{-6}$ \\
    $0.05$ & $1.73$ & $1.39$ & $256^2\times384$ & $4.8^{-0.6}_{+0.7}\times10^{-1}$ & $0.047\pm0.005$ & $1.4^{-0.3}_{+0.3}\times10^{-4}$ & $1.3^{-0.3}_{+0.5}\times10^{-6}$ \\
    $0.05$ & $0.55$ & $0.44$ & $256^2\times384$ & $1.6^{-0.4}_{+0.5}\times10^{-1}$ & $0.050\pm0.008$ & $1.3^{-0.2}_{+0.2}\times10^{-4}$ & $1.1^{-0.2}_{+0.3}\times10^{-6}$ \\
    $0.05$ & $21.33$ & $21.33$ & $256^2\times384$ & $1.1^{-0.3}_{+0.4}\times10^{1}$ & $0.053\pm0.005$ & $2.1^{-0.7}_{+1.1}\times10^{-6}$ & $2.2^{-0.5}_{+0.6}\times10^{-6}$ \\
    $0.05$ & $6.39$ & $6.39$ & $256^2\times384$ & $3.0^{-0.7}_{+0.8}\times10^{0}$ & $0.053\pm0.005$ & $1.5^{-0.7}_{+1.4}\times10^{-5}$ & $2.2^{-0.4}_{+0.5}\times10^{-6}$ \\
    $0.05$ & $3.00$ & $3.00$ & $256^2\times384$ & $9.8^{-1.8}_{+2.2}\times10^{-1}$ & $0.051\pm0.005$ & $1.2^{-0.4}_{+0.6}\times10^{-4}$ & $1.8^{-0.4}_{+0.5}\times10^{-6}$ \\
    $0.05$ & $1.96$ & $1.96$ & $256^2\times384$ & $4.9^{-0.7}_{+0.8}\times10^{-1}$ & $0.048\pm0.005$ & $2.5^{-0.5}_{+0.6}\times10^{-4}$ & $1.2^{-0.3}_{+0.5}\times10^{-6}$ \\
    $0.05$ & $0.62$ & $0.62$ & $512^2\times768$ & $1.6^{-0.4}_{+0.5}\times10^{-1}$ & $0.050\pm0.008$ & $2.5^{-0.4}_{+0.4}\times10^{-4}$ & $1.0^{-0.3}_{+0.3}\times10^{-6}$ \\
    $0.05$ & $0.62$ & $0.62$ & $1024^2\times1536$ & $1.6^{-0.2}_{+0.2}\times10^{-1}$ & $0.050\pm0.004$ & $2.4^{-0.4}_{+0.4}\times10^{-4}$ & $1.1^{-0.2}_{+0.3}\times10^{-6}$ \\
    $0.05$ & $0.28$ & $0.28$ & $512^2\times768$ & $7.2^{-1.1}_{+1.3}\times10^{-2}$ & $0.057\pm0.010$ & $1.9^{-0.3}_{+0.3}\times10^{-4}$ & $2.0^{-0.5}_{+0.7}\times10^{-6}$ \\
    $0.05$ & $21.87$ & $26.25$ & $256^2\times384$ & $1.1^{-0.2}_{+0.3}\times10^{1}$ & $0.053\pm0.005$ & $2.5^{-1.0}_{+1.8}\times10^{-6}$ & $2.1^{-0.5}_{+0.6}\times10^{-6}$ \\
    $0.05$ & $7.14$ & $8.57$ & $256^2\times384$ & $3.1^{-0.7}_{+0.9}\times10^{0}$ & $0.053\pm0.005$ & $1.7^{-0.8}_{+1.6}\times10^{-5}$ & $2.1^{-0.4}_{+0.6}\times10^{-6}$ \\
    $0.05$ & $3.00$ & $3.60$ & $256^2\times384$ & $9.5^{-1.7}_{+2.1}\times10^{-1}$ & $0.051\pm0.005$ & $2.0^{-0.6}_{+0.9}\times10^{-4}$ & $1.9^{-0.4}_{+0.4}\times10^{-6}$ \\
    $0.05$ & $2.00$ & $2.40$ & $256^2\times384$ & $5.0^{-0.7}_{+0.8}\times10^{-1}$ & $0.048\pm0.005$ & $3.5^{-0.7}_{+0.9}\times10^{-4}$ & $1.3^{-0.3}_{+0.5}\times10^{-6}$ \\
    $0.05$ & $0.60$ & $0.72$ & $512^2\times768$ & $1.5^{-0.4}_{+0.6}\times10^{-1}$ & $0.051\pm0.008$ & $3.7^{-0.6}_{+0.7}\times10^{-4}$ & $1.1^{-0.2}_{+0.3}\times10^{-6}$ \\
    $0.05$ & $0.20$ & $0.24$ & $512^2\times768$ & $5.2^{-0.8}_{+0.9}\times10^{-2}$ & $0.061\pm0.011$ & $3.6^{-0.5}_{+0.6}\times10^{-4}$ & $4.4^{-1.1}_{+1.4}\times10^{-6}$ \\
    $0.05$ & $17.78$ & $26.67$ & $256^2\times384$ & $8.8^{-2.2}_{+2.9}\times10^{0}$ & $0.053\pm0.005$ & $4.2^{-1.9}_{+3.7}\times10^{-6}$ & $2.1^{-0.5}_{+0.6}\times10^{-6}$ \\
    $0.05$ & $7.14$ & $10.71$ & $256^2\times384$ & $3.2^{-0.7}_{+0.9}\times10^{0}$ & $0.053\pm0.005$ & $2.5^{-1.2}_{+2.5}\times10^{-5}$ & $2.1^{-0.4}_{+0.6}\times10^{-6}$ \\
    $0.05$ & $3.20$ & $4.80$ & $256^2\times384$ & $1.1^{-0.2}_{+0.3}\times10^{0}$ & $0.051\pm0.005$ & $2.3^{-0.8}_{+1.2}\times10^{-4}$ & $1.9^{-0.4}_{+0.6}\times10^{-6}$ \\
    $0.05$ & $2.00$ & $3.00$ & $256^2\times384$ & $5.0^{-0.6}_{+0.7}\times10^{-1}$ & $0.047\pm0.004$ & $4.8^{-1.0}_{+1.2}\times10^{-4}$ & $1.2^{-0.3}_{+0.4}\times10^{-6}$ \\
    $0.05$ & $0.62$ & $0.93$ & $512^2\times768$ & $1.6^{-0.4}_{+0.6}\times10^{-1}$ & $0.050\pm0.008$ & $5.8^{-0.9}_{+1.1}\times10^{-4}$ & $1.1^{-0.3}_{+0.3}\times10^{-6}$ \\
    $0.05$ & $0.20$ & $0.30$ & $512^2\times768$ & $5.5^{-0.8}_{+1.0}\times10^{-2}$ & $0.063\pm0.011$ & $5.8^{-0.8}_{+0.9}\times10^{-4}$ & $3.8^{-0.8}_{+1.0}\times10^{-6}$ \\
    $0.05$ & $20.00$ & $39.99$ & $256^2\times384$ & $1.0^{-0.2}_{+0.3}\times10^{1}$ & $0.053\pm0.005$ & $4.1^{-1.8}_{+3.3}\times10^{-6}$ & $2.1^{-0.4}_{+0.5}\times10^{-6}$ \\
    $0.05$ & $6.67$ & $13.33$ & $256^2\times384$ & $3.1^{-0.7}_{+0.9}\times10^{0}$ & $0.052\pm0.005$ & $4.3^{-2.0}_{+3.6}\times10^{-5}$ & $2.0^{-0.5}_{+0.6}\times10^{-6}$ \\
    $0.05$ & $2.75$ & $5.50$ & $256^2\times384$ & $8.8^{-1.4}_{+1.7}\times10^{-1}$ & $0.050\pm0.005$ & $4.5^{-1.3}_{+1.8}\times10^{-4}$ & $1.8^{-0.4}_{+0.5}\times10^{-6}$ \\
    $0.05$ & $1.25$ & $2.50$ & $256^2\times384$ & $3.1^{-0.3}_{+0.4}\times10^{-1}$ & $0.046\pm0.005$ & $9.6^{-1.6}_{+1.9}\times10^{-4}$ & $1.1^{-0.3}_{+0.4}\times10^{-6}$ \\
    $0.05$ & $0.58$ & $1.16$ & $512^2\times768$ & $1.5^{-0.5}_{+0.7}\times10^{-1}$ & $0.050\pm0.008$ & $9.2^{-1.5}_{+1.8}\times10^{-4}$ & $1.2^{-0.3}_{+0.4}\times10^{-6}$ \\
    $0.05$ & $0.58$ & $1.16$ & $1024^2\times1536$ & $1.5^{-0.2}_{+0.2}\times10^{-1}$ & $0.051\pm0.007$ & $9.2^{-1.3}_{+1.7}\times10^{-4}$ & $1.2^{-0.3}_{+0.4}\times10^{-6}$ \\
    $0.05$ & $0.20$ & $0.40$ & $512^2\times768$ & $5.9^{-1.0}_{+1.3}\times10^{-2}$ & $0.064\pm0.011$ & $9.2^{-1.2}_{+1.4}\times10^{-4}$ & $4.3^{-0.9}_{+1.1}\times10^{-6}$ \\
    \hline
    $0.10$ & $8.66$ & $6.93$ & $256^2\times384$ & $9.5^{-2.3}_{+3.1}\times10^{0}$ & $0.104\pm0.012$ & $1.7^{-0.4}_{+0.5}\times10^{-5}$ & $3.0^{-0.7}_{+1.0}\times10^{-5}$ \\
    $0.10$ & $2.74$ & $2.19$ & $256^2\times384$ & $2.6^{-0.6}_{+0.8}\times10^{0}$ & $0.103\pm0.011$ & $5.0^{-2.0}_{+3.3}\times10^{-5}$ & $2.8^{-0.7}_{+0.9}\times10^{-5}$ \\
    $0.10$ & $0.87$ & $0.69$ & $256^2\times384$ & $4.9^{-0.7}_{+0.8}\times10^{-1}$ & $0.096\pm0.009$ & $5.6^{-1.0}_{+1.2}\times10^{-4}$ & $2.2^{-0.5}_{+0.7}\times10^{-5}$ \\
    $0.10$ & $0.27$ & $0.22$ & $512^2\times768$ & $1.6^{-0.3}_{+0.3}\times10^{-1}$ & $0.102\pm0.017$ & $6.0^{-1.0}_{+1.2}\times10^{-4}$ & $2.6^{-0.6}_{+0.8}\times10^{-5}$ \\
    $0.10$ & $9.59$ & $9.59$ & $256^2\times384$ & $9.8^{-2.3}_{+3.0}\times10^{0}$ & $0.105\pm0.010$ & $2.1^{-0.6}_{+0.8}\times10^{-5}$ & $3.2^{-0.7}_{+0.9}\times10^{-5}$ \\
    $0.10$ & $3.75$ & $3.75$ & $256^2\times384$ & $3.4^{-0.8}_{+1.0}\times10^{0}$ & $0.105\pm0.008$ & $5.6^{-2.1}_{+3.4}\times10^{-5}$ & $3.1^{-0.5}_{+0.7}\times10^{-5}$ \\
    $0.10$ & $1.56$ & $1.56$ & $256^2\times384$ & $1.0^{-0.2}_{+0.2}\times10^{0}$ & $0.100\pm0.010$ & $4.6^{-1.4}_{+2.0}\times10^{-4}$ & $2.7^{-0.6}_{+0.8}\times10^{-5}$ \\
    $0.10$ & $0.98$ & $0.98$ & $256^2\times384$ & $5.0^{-0.7}_{+0.8}\times10^{-1}$ & $0.094\pm0.009$ & $1.0^{-0.2}_{+0.3}\times10^{-3}$ & $2.0^{-0.5}_{+0.6}\times10^{-5}$ \\
    $0.10$ & $0.22$ & $0.22$ & $512^2\times768$ & $1.2^{-0.2}_{+0.2}\times10^{-1}$ & $0.114\pm0.021$ & $1.1^{-0.2}_{+0.2}\times10^{-3}$ & $3.6^{-1.0}_{+1.4}\times10^{-5}$ \\
    $0.10$ & $11.18$ & $13.41$ & $256^2\times384$ & $1.1^{-0.3}_{+0.4}\times10^{1}$ & $0.105\pm0.011$ & $2.1^{-0.6}_{+0.8}\times10^{-5}$ & $3.2^{-0.7}_{+1.0}\times10^{-5}$ \\
    $0.10$ & $3.76$ & $4.51$ & $256^2\times384$ & $3.2^{-0.7}_{+0.9}\times10^{0}$ & $0.105\pm0.011$ & $8.3^{-3.2}_{+5.3}\times10^{-5}$ & $3.3^{-0.8}_{+1.0}\times10^{-5}$ \\
    $0.10$ & $1.50$ & $1.80$ & $256^2\times384$ & $9.0^{-1.5}_{+1.8}\times10^{-1}$ & $0.100\pm0.008$ & $7.8^{-2.2}_{+3.0}\times10^{-4}$ & $2.9^{-0.6}_{+0.7}\times10^{-5}$ \\
    $0.10$ & $1.00$ & $1.20$ & $256^2\times384$ & $5.0^{-0.7}_{+0.8}\times10^{-1}$ & $0.094\pm0.008$ & $1.4^{-0.3}_{+0.3}\times10^{-3}$ & $1.9^{-0.4}_{+0.6}\times10^{-5}$ \\
    $0.10$ & $0.22$ & $0.26$ & $512^2\times768$ & $1.2^{-0.2}_{+0.2}\times10^{-1}$ & $0.114\pm0.020$ & $1.4^{-0.2}_{+0.3}\times10^{-3}$ & $2.8^{-0.7}_{+0.9}\times10^{-5}$ \\
    $0.10$ & $10.61$ & $15.92$ & $256^2\times384$ & $1.1^{-0.3}_{+0.4}\times10^{1}$ & $0.105\pm0.010$ & $2.4^{-0.8}_{+1.2}\times10^{-5}$ & $3.1^{-0.7}_{+0.8}\times10^{-5}$ \\
    $0.10$ & $3.75$ & $5.62$ & $256^2\times384$ & $3.3^{-0.7}_{+0.8}\times10^{0}$ & $0.104\pm0.009$ & $1.0^{-0.4}_{+0.8}\times10^{-4}$ & $3.0^{-0.6}_{+0.7}\times10^{-5}$ \\
    $0.10$ & $1.50$ & $2.25$ & $256^2\times384$ & $9.1^{-1.4}_{+1.6}\times10^{-1}$ & $0.100\pm0.007$ & $1.0^{-0.3}_{+0.4}\times10^{-3}$ & $3.0^{-0.5}_{+0.7}\times10^{-5}$ \\
    $0.10$ & $0.98$ & $1.47$ & $256^2\times384$ & $4.8^{-0.5}_{+0.6}\times10^{-1}$ & $0.096\pm0.008$ & $2.2^{-0.4}_{+0.5}\times10^{-3}$ & $2.0^{-0.5}_{+0.7}\times10^{-5}$ \\
    $0.10$ & $0.22$ & $0.33$ & $512^2\times768$ & $1.2^{-0.2}_{+0.3}\times10^{-1}$ & $0.114\pm0.021$ & $2.5^{-0.4}_{+0.4}\times10^{-3}$ & $3.2^{-0.8}_{+1.1}\times10^{-5}$ \\
    $0.10$ & $10.00$ & $20.00$ & $256^2\times384$ & $1.1^{-0.3}_{+0.4}\times10^{1}$ & $0.105\pm0.010$ & $2.8^{-1.0}_{+1.5}\times10^{-5}$ & $3.2^{-0.7}_{+0.9}\times10^{-5}$ \\
    $0.10$ & $3.50$ & $7.00$ & $256^2\times384$ & $3.3^{-0.7}_{+0.9}\times10^{0}$ & $0.104\pm0.009$ & $1.6^{-0.7}_{+1.3}\times10^{-4}$ & $3.1^{-0.7}_{+0.9}\times10^{-5}$ \\
    $0.10$ & $1.50$ & $3.00$ & $256^2\times384$ & $9.7^{-1.6}_{+1.9}\times10^{-1}$ & $0.099\pm0.009$ & $1.5^{-0.4}_{+0.6}\times10^{-3}$ & $3.1^{-0.7}_{+0.9}\times10^{-5}$ \\
    $0.10$ & $0.92$ & $1.83$ & $256^2\times384$ & $4.7^{-0.5}_{+0.5}\times10^{-1}$ & $0.097\pm0.008$ & $3.2^{-0.5}_{+0.6}\times10^{-3}$ & $2.2^{-0.5}_{+0.7}\times10^{-5}$ \\
    $0.10$ & $0.20$ & $0.40$ & $512^2\times768$ & $1.1^{-0.2}_{+0.3}\times10^{-1}$ & $0.114\pm0.019$ & $4.7^{-0.7}_{+0.8}\times10^{-3}$ & $5.1^{-1.3}_{+1.7}\times10^{-5}$ \\
    \hline
    $0.25$ & $3.46$ & $2.77$ & $256^2\times384$ & $8.9^{-1.9}_{+2.5}\times10^{0}$ & $0.244\pm0.025$ & $4.3^{-0.9}_{+1.2}\times10^{-4}$ & $8.8^{-2.0}_{+2.6}\times10^{-4}$ \\
    $0.25$ & $1.10$ & $0.88$ & $256^2\times384$ & $2.4^{-0.6}_{+0.7}\times10^{0}$ & $0.244\pm0.024$ & $8.5^{-2.0}_{+2.6}\times10^{-4}$ & $8.9^{-1.7}_{+2.1}\times10^{-4}$ \\
    $0.25$ & $0.35$ & $0.28$ & $256^2\times384$ & $5.1^{-1.6}_{+2.4}\times10^{-1}$ & $0.238\pm0.024$ & $4.2^{-0.7}_{+0.9}\times10^{-3}$ & $1.1^{-0.3}_{+0.4}\times10^{-3}$ \\
    $0.25$ & $0.12$ & $0.10$ & $512^2\times768$ & $1.6^{-0.2}_{+0.2}\times10^{-1}$ & $0.206\pm0.034$ & $2.5^{-0.4}_{+0.5}\times10^{-3}$ & $5.3^{-1.5}_{+2.0}\times10^{-4}$ \\
    $0.25$ & $6.40$ & $6.40$ & $256^2\times384$ & $1.1^{-0.2}_{+0.3}\times10^{1}$ & $0.248\pm0.023$ & $5.0^{-0.8}_{+0.9}\times10^{-4}$ & $9.8^{-1.8}_{+2.3}\times10^{-4}$ \\
    $0.25$ & $2.00$ & $2.00$ & $256^2\times384$ & $3.2^{-0.6}_{+0.7}\times10^{0}$ & $0.250\pm0.028$ & $1.1^{-0.2}_{+0.3}\times10^{-3}$ & $1.1^{-0.2}_{+0.3}\times10^{-3}$ \\
    $0.25$ & $0.75$ & $0.75$ & $256^2\times384$ & $1.2^{-0.2}_{+0.3}\times10^{0}$ & $0.236\pm0.023$ & $2.4^{-0.6}_{+0.9}\times10^{-3}$ & $8.0^{-1.8}_{+2.4}\times10^{-4}$ \\
    $0.25$ & $0.39$ & $0.39$ & $256^2\times384$ & $5.1^{-2.1}_{+4.5}\times10^{-1}$ & $0.233\pm0.021$ & $6.6^{-1.4}_{+1.8}\times10^{-3}$ & $8.1^{-1.8}_{+2.3}\times10^{-4}$ \\
    $0.25$ & $0.12$ & $0.12$ & $512^2\times768$ & $1.8^{-0.2}_{+0.2}\times10^{-1}$ & $0.266\pm0.043$ & $8.5^{-1.5}_{+1.9}\times10^{-3}$ & $1.2^{-0.3}_{+0.3}\times10^{-3}$ \\
    $0.25$ & $0.12$ & $0.12$ & $1024^2\times1536$ & $1.8^{-0.2}_{+0.2}\times10^{-1}$ & $0.267\pm0.009$ & $8.5^{-1.4}_{+1.6}\times10^{-3}$ & $1.2^{-0.2}_{+0.3}\times10^{-3}$ \\
    $0.25$ & $7.00$ & $8.40$ & $256^2\times384$ & $1.2^{-0.2}_{+0.2}\times10^{1}$ & $0.248\pm0.020$ & $5.1^{-0.8}_{+0.9}\times10^{-4}$ & $9.7^{-1.7}_{+2.1}\times10^{-4}$ \\
    $0.25$ & $2.00$ & $2.40$ & $256^2\times384$ & $3.1^{-0.5}_{+0.7}\times10^{0}$ & $0.248\pm0.023$ & $1.4^{-0.3}_{+0.3}\times10^{-3}$ & $1.0^{-0.2}_{+0.2}\times10^{-3}$ \\
    $0.25$ & $0.75$ & $0.90$ & $256^2\times384$ & $1.1^{-0.2}_{+0.3}\times10^{0}$ & $0.237\pm0.024$ & $3.5^{-0.9}_{+1.3}\times10^{-3}$ & $8.4^{-2.0}_{+2.7}\times10^{-4}$ \\
    $0.25$ & $0.40$ & $0.48$ & $256^2\times384$ & $5.1^{-2.1}_{+4.1}\times10^{-1}$ & $0.232\pm0.021$ & $9.3^{-1.8}_{+2.5}\times10^{-3}$ & $7.7^{-1.9}_{+2.6}\times10^{-4}$ \\
    $0.25$ & $0.10$ & $0.12$ & $512^2\times768$ & $1.9^{-0.2}_{+0.2}\times10^{-1}$ & $0.362\pm0.043$ & $2.3^{-0.4}_{+0.5}\times10^{-2}$ & $4.4^{-1.0}_{+1.4}\times10^{-3}$ \\
    $0.25$ & $6.40$ & $9.60$ & $256^2\times384$ & $1.1^{-0.2}_{+0.3}\times10^{1}$ & $0.252\pm0.022$ & $5.9^{-1.0}_{+1.2}\times10^{-4}$ & $1.0^{-0.2}_{+0.2}\times10^{-3}$ \\
    $0.25$ & $2.00$ & $3.00$ & $256^2\times384$ & $3.2^{-0.5}_{+0.7}\times10^{0}$ & $0.249\pm0.024$ & $1.6^{-0.4}_{+0.6}\times10^{-3}$ & $9.9^{-1.8}_{+2.2}\times10^{-4}$ \\
    $0.25$ & $0.82$ & $1.24$ & $256^2\times384$ & $1.4^{-0.3}_{+0.3}\times10^{0}$ & $0.236\pm0.025$ & $3.6^{-1.0}_{+1.4}\times10^{-3}$ & $8.4^{-2.2}_{+3.0}\times10^{-4}$ \\
    $0.25$ & $0.39$ & $0.59$ & $256^2\times384$ & $4.9^{-1.8}_{+3.2}\times10^{-1}$ & $0.231\pm0.023$ & $1.3^{-0.3}_{+0.4}\times10^{-2}$ & $7.6^{-1.7}_{+2.2}\times10^{-4}$ \\
    $0.25$ & $0.10$ & $0.15$ & $512^2\times768$ & $2.2^{-0.2}_{+0.3}\times10^{-1}$ & $0.449\pm0.053$ & $6.7^{-1.1}_{+1.4}\times10^{-2}$ & $1.2^{-0.3}_{+0.3}\times10^{-2}$ \\
    $0.25$ & $6.00$ & $12.00$ & $256^2\times384$ & $1.1^{-0.2}_{+0.2}\times10^{1}$ & $0.251\pm0.023$ & $6.4^{-1.1}_{+1.3}\times10^{-4}$ & $1.0^{-0.2}_{+0.2}\times10^{-3}$ \\
    $0.25$ & $2.00$ & $4.00$ & $256^2\times384$ & $3.3^{-0.5}_{+0.6}\times10^{0}$ & $0.247\pm0.021$ & $2.0^{-0.5}_{+0.7}\times10^{-3}$ & $9.7^{-1.6}_{+2.0}\times10^{-4}$ \\
    $0.25$ & $0.75$ & $1.50$ & $256^2\times384$ & $1.3^{-0.2}_{+0.3}\times10^{0}$ & $0.235\pm0.024$ & $5.9^{-1.6}_{+2.3}\times10^{-3}$ & $7.8^{-2.1}_{+2.8}\times10^{-4}$ \\
    $0.25$ & $0.37$ & $0.73$ & $256^2\times384$ & $5.2^{-1.4}_{+2.0}\times10^{-1}$ & $0.232\pm0.022$ & $1.7^{-0.4}_{+0.5}\times10^{-2}$ & $8.0^{-1.7}_{+2.2}\times10^{-4}$ \\
    $0.25$ & $0.10$ & $0.20$ & $512^2\times768$ & $2.9^{-0.5}_{+0.7}\times10^{-1}$ & $0.536\pm0.070$ & $1.4^{-0.3}_{+0.3}\times10^{-1}$ & $3.0^{-0.7}_{+0.9}\times10^{-2}$ \\
    $0.25$ & $0.10$ & $0.20$ & $1024^2\times1536$ & $2.9^{-0.3}_{+0.3}\times10^{-1}$ & $0.550\pm0.083$ & $1.4^{-0.3}_{+0.4}\times10^{-1}$ & $3.2^{-0.7}_{+0.9}\times10^{-2}$ \\
    \hline
    $0.40$ & $1.75$ & $1.75$ & $512^2\times768$ & $7.4^{-1.4}_{+1.7}\times10^{0}$ & $0.365\pm0.069$ & $2.4^{-0.9}_{+1.5}\times10^{-3}$ & $4.3^{-1.8}_{+3.2}\times10^{-3}$ \\
    $0.40$ & $0.70$ & $0.70$ & $512^2\times768$ & $2.7^{-0.6}_{+0.8}\times10^{0}$ & $0.371\pm0.065$ & $4.9^{-1.6}_{+2.5}\times10^{-3}$ & $5.4^{-2.0}_{+3.2}\times10^{-3}$ \\
    $0.40$ & $0.30$ & $0.30$ & $512^2\times768$ & $7.3^{-2.5}_{+4.7}\times10^{-1}$ & $0.400\pm0.067$ & $2.2^{-0.4}_{+0.6}\times10^{-2}$ & $1.3^{-0.3}_{+0.5}\times10^{-2}$ \\
    $0.40$ & $0.15$ & $0.15$ & $512^2\times768$ & $3.3^{-0.4}_{+0.5}\times10^{-1}$ & $0.402\pm0.079$ & $3.2^{-0.8}_{+1.0}\times10^{-2}$ & $1.1^{-0.3}_{+0.4}\times10^{-2}$ \\
    $0.40$ & $1.75$ & $2.10$ & $512^2\times768$ & $7.8^{-1.4}_{+1.7}\times10^{0}$ & $0.367\pm0.071$ & $2.6^{-1.0}_{+1.8}\times10^{-3}$ & $4.4^{-1.8}_{+3.2}\times10^{-3}$ \\
    $0.40$ & $0.70$ & $0.84$ & $512^2\times768$ & $2.8^{-0.6}_{+0.8}\times10^{0}$ & $0.378\pm0.065$ & $5.5^{-2.0}_{+3.3}\times10^{-3}$ & $5.9^{-2.1}_{+3.4}\times10^{-3}$ \\
    $0.40$ & $0.30$ & $0.36$ & $512^2\times768$ & $7.7^{-2.3}_{+4.0}\times10^{-1}$ & $0.402\pm0.068$ & $2.3^{-0.5}_{+0.7}\times10^{-2}$ & $1.2^{-0.3}_{+0.5}\times10^{-2}$ \\
    $0.40$ & $0.17$ & $0.20$ & $512^2\times768$ & $4.2^{-0.7}_{+0.9}\times10^{-1}$ & $0.436\pm0.090$ & $4.5^{-1.0}_{+1.3}\times10^{-2}$ & $1.7^{-0.5}_{+0.7}\times10^{-2}$ \\
    $0.40$ & $1.37$ & $2.06$ & $512^2\times768$ & $6.8^{-1.3}_{+1.7}\times10^{0}$ & $0.365\pm0.075$ & $2.9^{-1.1}_{+1.7}\times10^{-3}$ & $4.3^{-1.9}_{+3.3}\times10^{-3}$ \\
    $0.40$ & $0.60$ & $0.90$ & $512^2\times768$ & $2.6^{-0.6}_{+0.7}\times10^{0}$ & $0.383\pm0.069$ & $7.1^{-2.6}_{+4.2}\times10^{-3}$ & $6.6^{-2.3}_{+3.7}\times10^{-3}$ \\
    $0.40$ & $0.30$ & $0.45$ & $512^2\times768$ & $9.0^{-2.2}_{+3.0}\times10^{-1}$ & $0.400\pm0.070$ & $2.5^{-0.6}_{+0.8}\times10^{-2}$ & $1.2^{-0.4}_{+0.6}\times10^{-2}$ \\
    $0.40$ & $0.14$ & $0.21$ & $512^2\times768$ & $4.7^{-0.8}_{+1.1}\times10^{-1}$ & $0.543\pm0.080$ & $9.0^{-1.5}_{+1.8}\times10^{-2}$ & $4.1^{-0.8}_{+1.1}\times10^{-2}$ \\
    $0.40$ & $1.33$ & $2.67$ & $512^2\times768$ & $7.3^{-1.4}_{+1.7}\times10^{0}$ & $0.367\pm0.070$ & $3.0^{-1.2}_{+2.0}\times10^{-3}$ & $4.7^{-2.0}_{+3.5}\times10^{-3}$ \\
    $0.40$ & $0.50$ & $1.00$ & $512^2\times768$ & $2.2^{-0.5}_{+0.6}\times10^{0}$ & $0.390\pm0.067$ & $1.1^{-0.3}_{+0.5}\times10^{-2}$ & $7.4^{-2.5}_{+4.0}\times10^{-3}$ \\
    $0.40$ & $0.27$ & $0.54$ & $512^2\times768$ & $9.5^{-2.3}_{+3.1}\times10^{-1}$ & $0.397\pm0.077$ & $3.3^{-0.9}_{+1.3}\times10^{-2}$ & $1.5^{-0.5}_{+0.8}\times10^{-2}$ \\
    $0.40$ & $0.12$ & $0.25$ & $512^2\times768$ & $7.1^{-3.5}_{+9.1}\times10^{-1}$ & $0.585\pm0.153$ & $1.0^{-0.3}_{+0.4}\times10^{-1}$ & $4.9^{-1.8}_{+2.9}\times10^{-2}$ \\
	\hline
			
\end{longtable}
\end{center}

%\twocolumn
%%%%%%%%%%%%%%%%%%%%%%%%%%%%%%%%%%%%%%%%%%%%%%%%%%
% Don't change these lines
\bsp	% typesetting comment
\label{lastpage}
\end{document}